\documentclass[aps,prb,preprint,superscriptaddress,a4paper,nofootinbib]{revtex4-2}

\usepackage[utf8]{inputenc}
\usepackage[]{hyperref}
\usepackage[]{graphicx}
\usepackage{miller}
\usepackage[]{amsmath}
\newcommand{\etal}{\textit{et al}.}
\newcommand{\ie}{\textit{i.e.}}
\newcommand{\cf}{\textit{cf}}
\newcommand{\insitu}{\textit{in situ}}
\newcommand{\Insitu}{\textit{In situ}}
\newcommand{\abinitio}{\textit{ab initio}}
\newcommand{\Abinitio}{\textit{Ab initio}}
\newcommand{\vasp}{\textsc{Vasp}}
\newcommand{\phonopy}{\textsc{Phonopy}}
\newcommand{\maud}{\textsc{Maud}}

\newcommand{\degree}{$^{\circ}$}

\newcommand{\xO}{x_{\rm O}}
\begin{document}

\title{Impact of oxygen ordering on titanium lattice parameters}

\author{Martin S. Talla Noutack}
\affiliation{Univ. Paris-Saclay, CEA, Service de recherche en Corrosion et Comportement des Matériaux,  SRMP,  F-91191 Gif-sur-Yvette, France}

\author{Fabienne Amann}
\affiliation{Univ. PSL, Chimie ParisTech, CNRS, IRCP (UMR 8247), F-75005, Paris, France}
\affiliation{Univ. Paris Est Creteil, CNRS, ICMPE (UMR 7182), 2 rue Henri Dunant, F-94320, Thiais, France}

\author{Sophie Nowak}
\affiliation{Univ. Paris Cité, CNRS, ITODYS (UMR 7086), F-75013, Paris, France}

\author{Régis Poulain}
\affiliation{Univ. Paris Est Creteil, CNRS, ICMPE (UMR 7182), 2 rue Henri Dunant, F-94320, Thiais, France}

\author{Raphaëlle Guillou}
\affiliation{Univ. Paris-Saclay, CEA, Service de Recherche sur les Matériaux et procédés Avancés, F-91191 Gif-sur-Yvette, France}

\author{Stéphanie Delannoy}
\affiliation{Univ. PSL, Chimie ParisTech, CNRS, IRCP (UMR 8247), F-75005, Paris, France}
\affiliation{Biotech Dental SAS, 305 allées de Craponne, F-13300, Salon-de-Provence, France}

\author{Ivan Guillot}
\affiliation{Univ. Paris Est Creteil, CNRS, ICMPE (UMR 7182), 2 rue Henri Dunant, F-94320, Thiais, France}

\author{Frédéric Prima}
\affiliation{Univ. PSL, Chimie ParisTech, CNRS, IRCP (UMR 8247), F-75005, Paris, France}

\author{Emmanuel Clouet}
\thanks{Corresponding author}
\email[]{emmanuel.clouet@cea.fr}
\affiliation{Univ. Paris-Saclay, CEA, Service de recherche en Corrosion et Comportement des Matériaux,  SRMP, F-91191 Gif-sur-Yvette, France}

\date{\today}

\begin{abstract}
	Variations with oxygen concentration of titanium lattice parameters are obtained by means
	of \abinitio{} calculations, considering the impact of oxygen ordering. 
	The quasi harmonic approximation is used to take into account the thermal expansion 
	at finite temperature. 
	Results show that lattice parameters depend mainly on oxygen concentration
	and to a lesser extent to the ordering state. 
	Knowing these theoretical variations, one can get insights on the composition 
	of ordered compounds existing in Ti-O binary alloys 
	from their lattice mismatch measured experimentally by X-ray diffraction.
	The approach is used in a binary alloy containing 6000\,ppm in weight of oxygen.
	It is concluded that the ordered compounds, which are observed after a recrystallization heat treatment, 
	do not have the expected Ti$_6$O stoichiometry but have a composition close 
	to the nominal concentration. 
	Oxygen ordering proceeds therefore before oxygen partitioning in titanium.
\end{abstract}

\maketitle
\clearpage

\section{Introduction}

Oxygen is one of the major hardening elements of titanium, used to strengthen the hexagonal close-packed (hcp) $\alpha$ phase \cite{Conrad1981}.
With a high solubility limit, more than 30\,at.\% at room temperature \cite{Murray1987}, 
titanium can incorporate a large amount of oxygen in its $\alpha$ matrix.
But suboxide compounds with a Ti$_6$O and a Ti$_3$O stoichiometry are known to exist below the solubility limit 
\cite{Kornilov1965,Yamaguchi1966,Yamaguchi1969}. 
These suboxides correspond to an ordering of the oxygen atoms which occupy the octahedral interstitial sites
of the hcp lattice \cite{Hirabayashi1974,Banerjee2007}.
\Abinitio{} calculations predict that these ordered compounds are the ground states in the titanium rich part 
of the Ti-O phase diagram \cite{Ruban2010,Burton2012c,Gunda2018}.
Although known for a long time now, one hardly considers the existence of these ordered compounds 
when discussing the effect of oxygen on titanium properties, in particular mechanical properties 
\cite{Yu2015,Barkia2015,Barkia2017,Chaari2019,Chong2020,Chong2023}. 
One generally assumes that all oxygen is dissolved in solid solution for the low concentrations 
encountered in titanium alloys which are usually lower than 1\,at.\% 
(0.5 and 1\,at.\%  respectively in grade-2 and -4 commercially pure Ti). 
But recent experiments have shown that ordered compounds 
can exist at room temperature in titanium for an oxygen concentration as low as 0.5\,at.\% \cite{Poulain2022},
with potentially important consequences on titanium mechanical properties \cite{Kornilov1973,Hirabayashi1974,Banerjee2007,Amann2023}.
These ordered compounds could be imaged by transmission electron microscopy 
through the superlattice reflections arising from oxygen ordering.
Lattice mismatches between the ordered precipitates and the $\alpha$ matrix
could also be measured at room temperature with X-ray diffraction (XRD). 
But no information on their composition could be obtained. 
In particular, it remains unclear if these ordered precipitates have a Ti$_6$O composition,
as expected from thermodynamics \cite{Gunda2018}, or if their oxygen content is lower, 
with oxygen ordering proceeding before oxygen partitioning\,
\ie{} with the formation of compounds having the same long-range order as Ti$_6$O ordered compounds
but without reaching the Ti$_6$O stoichiometry. 
As oxygen addition is known to increase the lattice parameters of the Ti $\alpha$-phase, 
one can attempt to deduce the composition of ordered compounds from the lattice mismatches measured by XRD. 
But one needs first to better know the impact of oxygen ordering on titanium lattice parameters,
in particular to be able to separate variations arising from oxygen ordering and from oxygen enrichment.

The aim of this article is to use \abinitio{} calculations to fully relate 
the variations of titanium lattice parameters to oxygen concentration and order state. 
We first perform \abinitio{} calculations at 0\,K and then extend them at finite temperature,
considering thermal expansion through the quasi-harmonic approximation. 
The theoretical variations obtained for lattice parameters are finally compared to the ones measured by XRD 
in a Ti-O binary alloy at different temperatures, allowing to conclude on the composition of the ordered 
compounds observed in this alloy.

\section{Phase stability and lattice parameters at 0 K}

\subsection{DFT parameters}

\Abinitio{} calculations are performed within non-spin polarized density functional theory (DFT) 
using the \vasp{} code \cite{Kresse1996}. 
The exchange and correlation functional is described with the generalized gradient approximation (GGA) as formulated by Perdew, Burke and Ernzerhof (PBE) \cite{Perdew1996}. 
Considering the local density approximation (LDA) instead of GGA leads to the same impact of oxygen
on titanium lattice parameters at 0\,K (appendix \ref{sec:LDA}).
Interactions between core and valence electrons are accounted for by the projector augmented wave (PAW) method \cite{Bloechl1994},
with $3s$, $3p$, $3d$, and $4s$ orbitals treated as valence states for Ti and $2s$ and $2p$ for O. 
A plane wave cutoff energy equal to 500\,eV is used for wave functions. 
A 24$\times$24$\times$18 $k$-point grid generated with the Monkhorst-Pack \cite{Monkhorst1976} method is used for primitive hcp Ti unit cell 
and grids with similar $k$-point density for supercells.
Integration is performed with the Methfessel-Paxton broadening scheme, using a 0.1\,eV width.
According to our convergence tests, this leads to a precision better than 1\,meV per atom on total energy differences.
Each structure is fully relaxed, \ie{} both periodicity vectors and atomic coordinates, to achieve a force convergence lower than 0.02\,eV/{\AA}  
and an energy convergence of 10$^{-6}$\,eV.
Although the supercells can theoretically become monoclinic during the relaxation,
the underlying lattices remain always close to the initial HCP crystal, with only a small orthorhombicity,
which may appear when the symmetry of the initial structure allows for different relaxations 
in different directions of the basal plane.

\subsection{Phase stability}

\begin{figure}[!bp]
	\centering
	\includegraphics[width=0.99\linewidth]{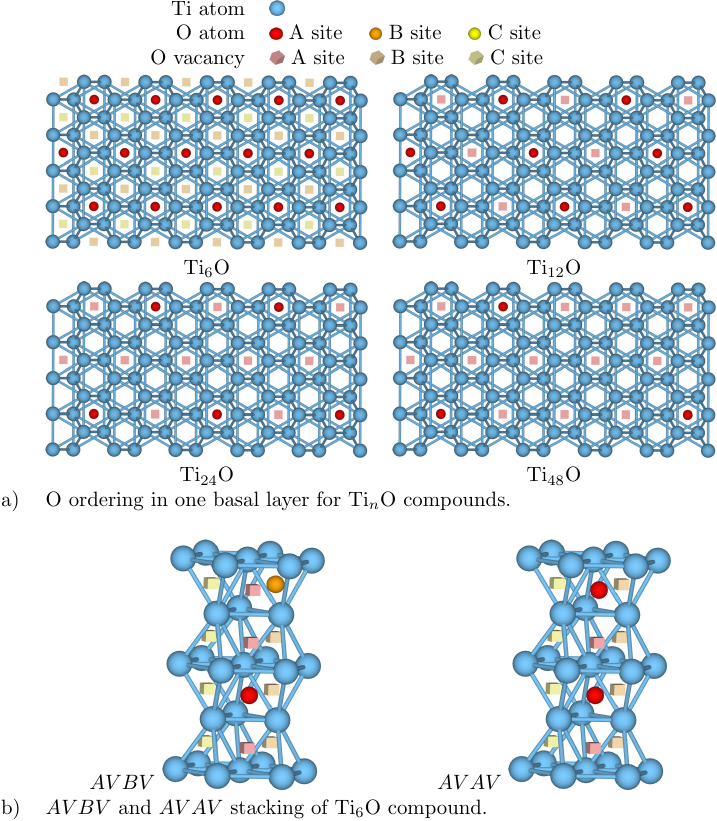}
	\caption{Structures of ordered Ti$_n$O compounds.
	(a) Projection in the basal plane showing the periodic occupation by O atoms 
	of octahedral interstitial sites for one basal layer.
	(b) $AVBV$ and (c) $AVAV$ stacking along the \hkl<c> axis of partially occupied basal layers.
	O atoms and empty octahedral sites are shown respectively with balls and cubes, 
	using a different color for $A$, $B$ and $C$ octahedral sites. 
	$B$ and $C$ empty octahedral sites are shown only for Ti$_6$O compound in (a).
	}
	\label{fig:TinO_structures}
\end{figure}

Oxygen occupies the octahedral interstitial sites of the hcp lattice formed by Ti atoms.
The ground state with the lowest oxygen content is known to have Ti$_6$O stoichiometry \cite{Hirabayashi1974,Banerjee2007,Ruban2010,Burton2012c,Gunda2018}. 
It corresponds to a stacking along the \hkl<c> axis of basal layers with one third of their octahedral sites 
occupied by oxygen atoms alternating with oxygen free layers.
Enriched basal layers are ordered with occupied octahedral sites separated by \hkl<1-100> periodicity vectors 
(see Ti$_6$O structure in Fig. \ref{fig:TinO_structures}a).  
Denoting $A$, $B$, and $C$ the three different octahedral sites in a basal layer, the Ti$_6$O ground state 
can be described by a periodic $AVBV$ stacking along the \hkl<c> axis, 
with $A$ corresponding to a basal layer where $A$ octahedral sites are occupied by O atoms
and $V$ meaning a layer where all octahedral sites are empty (Fig. \ref{fig:TinO_structures}b).

We check that our \abinitio{} approach predicts the right ground state in the concentration range 
$0\leq x_{\rm O} \leq 1/6$, where $x_{\rm O} = N_{\rm O} / N_{\rm Ti}$ is the occupation of the octahedral interstitial sites
\footnote{One can also define the oxygen concentration, $c_{\rm O} = N_{\rm O} /( N_{\rm Ti} + N_{\rm O} )$,
\ie{} the ratio of the number of oxygen atoms to the total number of atoms.}.
To this aim we calculate the formation energy of different ordered and disordered compounds. 
Disordered structures are built from a supercell corresponding to the $4\times4\times3$ repetition of the conventional hcp unit cell, 
with 96 Ti atoms.  Oxygen atoms are placed on the octahedral sites with the special quasirandom structures (SQS) method \cite{vandeWalle2013}
to obtain a random Ti-O solid solution.
Ordered Ti$_n$O structures with $n=6, 12, 24, 48$ are built in the same supercell, 
starting from the Ti$_6$O ordered compound and depleting oxygen atoms regularly in the basal layer, 
with either a $AVBV$ or $AVAV$ stacking along the \hkl<c> axis (Fig. \ref{fig:TinO_structures}). 
Finally, some partially ordered structures are also considered, 
using a larger supercell with 216 Ti atoms and incorporating some oxygen vacancies in the 
ordered Ti$_6$O compound.

\begin{figure}[!b]
	\centering
	\includegraphics[width=0.7\linewidth]{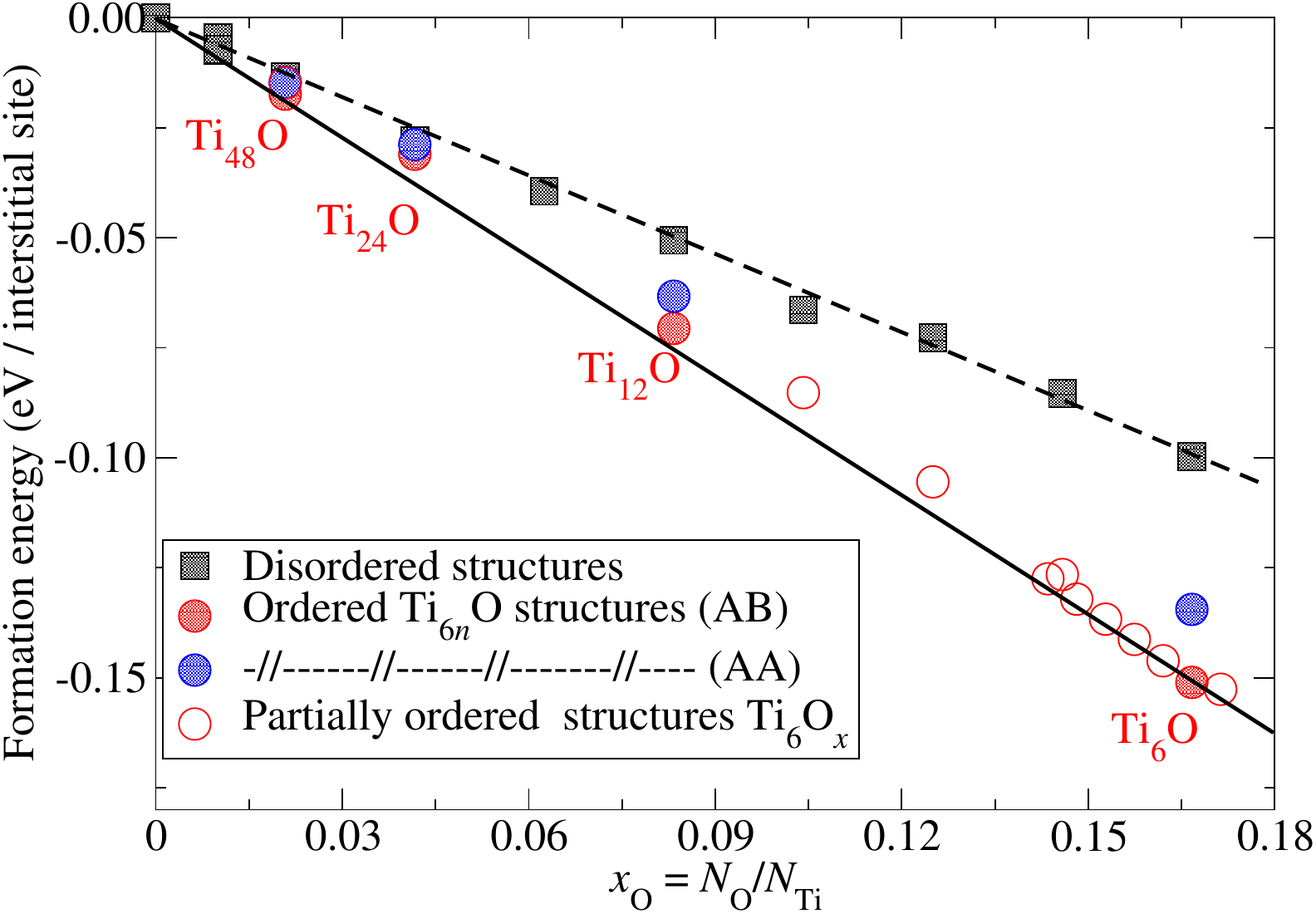}
	\caption{Formation energies of different Ti$_n$O$_m$ compounds as a function of
	oxygen occupation $x_{\rm O}=m/n$ of interstitial sites.
	Symbols correspond to \abinitio{} results for disordered, partly ordered
	or completely ordered compounds. 
	The solid black line indicates the convex hull linking most stable structures.
	}
	\label{fig:formationE}
\end{figure}

Formation energies are calculated with reference to pure Ti and to TiO compound, 
\ie{} structures with all octahedral sites either empty ($x_{\rm O}=0$) 
or occupied by oxygen atoms ($x_{\rm O}=1$).
Considering that the alloy is actually an oxygen-vacancy binary system 
on the octahedral interstitial sites of the hcp lattice formed by the Ti atoms, 
all formation energies are defined per interstitial site, leading to 
\begin{equation*}
	E^{\rm f}[{\rm Ti}_n{\rm O}_m] = \frac{1}{n} \left( 
	E[{\rm Ti}_n{\rm O}_m] - \frac{n-m}{n} E[{\rm Ti}_n] - \frac{m}{n} E[{\rm Ti}_n{\rm O}_n] \right),
\end{equation*}
where $E[{\rm Ti}_n]$ and $E[{\rm Ti}_n{\rm O}_n]$ are the energy of pure Ti and TiO in the same supercell 
as the Ti$_n$O$_m$ compound.

Fig. \ref{fig:formationE} shows that ordered structures are always more stable than disordered solid solutions,
confirming the ordering tendency of oxygen in titanium as seen experimentally
\cite{Yamaguchi1966,Yamaguchi1969,Yamaguchi1970,Hirabayashi1974,Poulain2022}.
No intermediate compound is found below the convex hull linking Ti and Ti$_6$O,
indicating that a supersaturated Ti(O) solid solution should decompose in an oxygen-depleted solution 
and an ordered Ti$_6$O compound at low temperature.

\begin{table}[t!]
\begin{center}
\caption{Energy variation $\Delta E$ of the Ti$_6$O structures
for different stackings of the oxygen enriched basal layers along the \hkl<c> direction.
}
\label{tab:Ti6O} 
\begin{tabular}{cc}
\hline
\hline
Stacking & $\Delta E$ (meV/site)\\
\hline
 $AVBV$     & $0.$\\
 $AVAV$     & $14.72$\\
 $AVBVCV$   & $-0.93$\\
 $AVBVCVBV$ & $-0.66$\\
\hline
\hline
\end{tabular}
\end{center}
\end{table}

Considering now the different possible stackings along the \hkl<c> axis of oxygen enriched basal layers,
ordered compounds with an $AVBV$ stacking are always more stable than $AVAV$ stacking, 
in agreement with previous \abinitio{} calculations \cite{Ruban2010,Burton2012c,Gunda2018} 
and with the stable structure known experimentally for Ti$_6$O.
Nevertheless, the stacking of these basal layers offers more possibilities. 
\Abinitio{} calculations for Ti$_6$O show that stackings in which two successive occupations
of oxygen layers are different, \ie{} $AVBV$, $AVBVCV$, $AVBVCVBV$, lead almost to the same formation energy,
with an energy difference lower than 1\,meV/site, thus below the expected DFT accuracy (Tab. \ref{tab:Ti6O}).
These ordered structures appear therefore degenerate, with $AVBV$ stacking 
having the smallest wave-length along \hkl<c> direction, leading to Devil's staircases \cite{Burton2012c}
disordering at finite temperature.

\subsection{Variations of lattice parameters}

\begin{figure}[!b]
	\centering
	\includegraphics[width=0.7\linewidth]{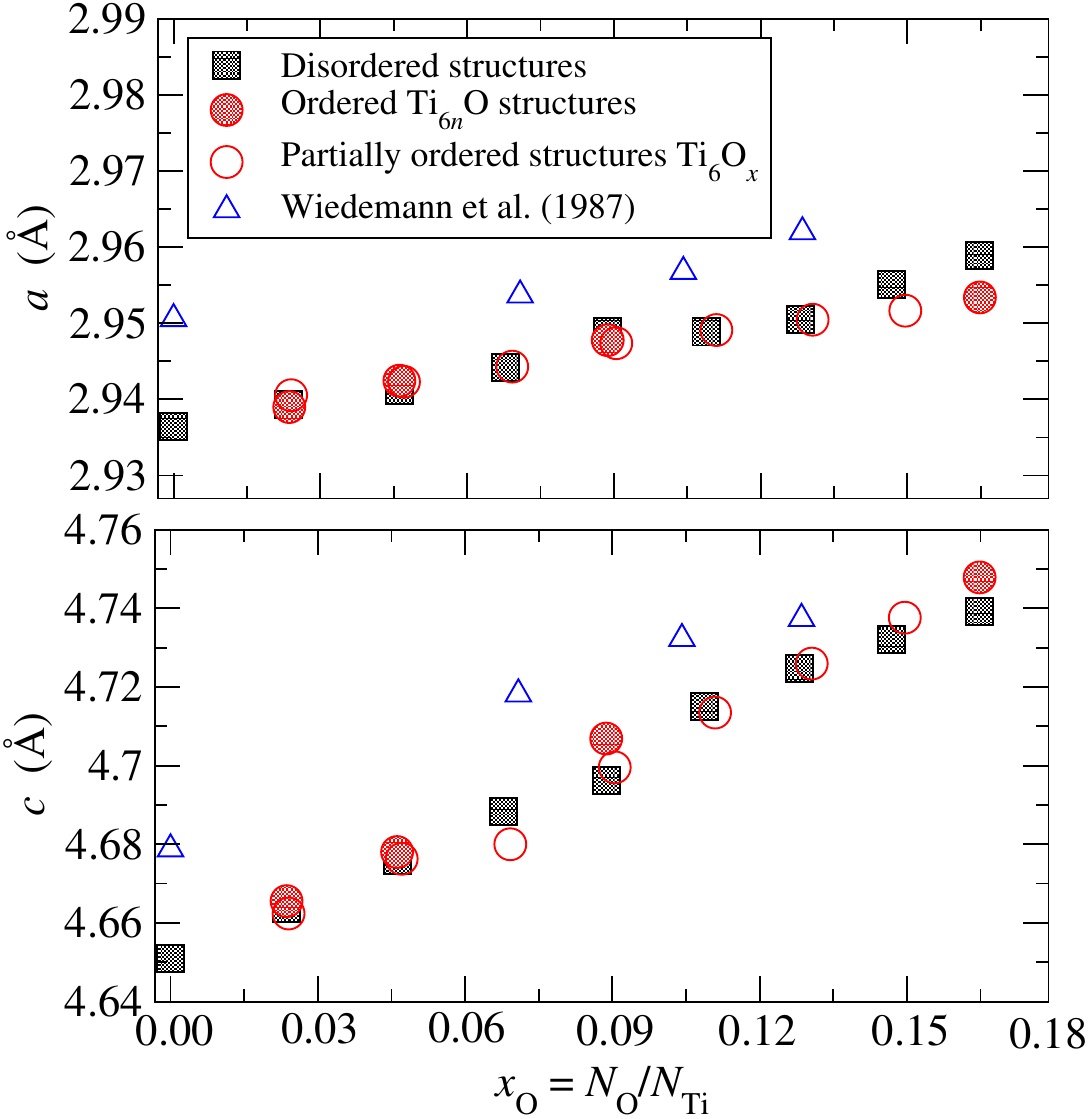}
	\caption{Lattice parameters determined at 0\,K by \abinitio{} calculations
	for different compounds of the Ti-O binary alloy (square and circle symbols)
	compared to the experimental values determined at room temperature by Wiedemann \cite{Wiedemann1987}.
	Similar experimental variations are shown in the review of Murray and Wriedt \cite{Murray1987}.
	}
	\label{fig:lattice_0K}
\end{figure}

\Abinitio{} calculations show that both $a$ and $c$ lattice parameters
increase with the oxygen content (Fig. \ref{fig:lattice_0K}),
with the parameter $c$ increasing more rapidly than $a$,
in agreement with experiments \cite{Murray1987,Wiedemann1987}. 
The comparison between experimental and theoretical values (Fig. \ref{fig:lattice_0K})
shows that \abinitio{} calculations correctly predict the increase rates of both lattice parameters
with the oxygen concentration, despite a slight underestimation of the absolute value.
As \abinitio{} calculations are performed at 0\,K while experimental data have been obtained at room temperature,
such an underestimation is indeed expected.  Besides, DFT approximation is known to be not fully quantitative
when predicting absolute values for lattice parameters, with a dependence on the functional 
used for the electronic exchange and correlation (see appendix \ref{sec:LDA}).

According to these \abinitio{} calculations, the lattice parameters mainly depend on the oxygen concentration,
with a marginal impact of the ordering state: for a same oxygen composition, a random solid solution 
and an ordered compound have almost the same lattice parameters at 0\,K (Fig. \ref{fig:lattice_0K}). 
The only noticeable impact is for the highest oxygen concentrations, in particular for Ti$_6$O, 
where larger $c$ and smaller $a$ parameters are obtained for ordered compounds 
than for disordered solid solutions.

\clearpage
\section{Modeling at finite temperatures}

\subsection{Methods}
\label{sec:finiteT:methods}

To go beyond 0\,K calculations, we now take into account thermal expansion 
by including the contribution of atomic vibrations in the formation free energy of the different compounds. 
Calculations are performed in the quasi-harmonic approximation, as described below, 
considering different structures in the same supercell containing 72 Ti atoms 
and defined by the periodicity vectors \hkl[-2200], \hkl[0-220], and \hkl[-1103].

This is done by coupling \phonopy{} \cite{Togo2015} with \vasp{} DFT calculations, so as to obtain phonons
for different volumes $V$ and ratios $\gamma=c/a$ of the supercell.
A finite displacement of 0.01\,\AA{} is used in \phonopy{} to calculate the supercell force constants.
Dynamical matrices $D(\vec{q})$ are then calculated by Fourier transformation on a regular $\vec{q}$ grid
with $N_q$ points and diagonalized to obtain the pulsations $\omega_{\vec{q},s}$ of the eigenmodes. 
The vibration energy, defined per interstitial octahedral site, of a supercell containing $n$ Ti atoms and $m$ O atoms is finally given by
\begin{equation}
	F^{\rm vib}(V,\gamma,T) = \frac{1}{n}\frac{1}{N_{\vec{q}}}
	    \sum_{\vec{q},s}{\left\{ 
		\frac{\hbar \omega_{\vec{q},s}(V,\gamma)}{2} 
		+ kT\ln{\left[1-\exp\left(-\frac{\hbar \omega_{\vec{q},s}(V,\gamma)}{kT}\right )\right]}
		\right\}}.
    \label{eq:Fvib}
\end{equation}
A $15\times15\times15$ $\vec{q}$ grid is used for Fourier transform.  
Comparison with results obtained with $10\times10\times10$ and $20\times20\times20$ grids
shows that this is enough for the supercell considered in this work.

\begin{table}[!t]
\begin{center}
	\caption{Contributions to the free energy of electronic excitations $F^{\rm el}$ 
	and of atomic vibrations $F^{\rm vib}$ calculated at different temperatures 
	in pure Ti and in ordered Ti$_6$O compound.
	Calculations are performed at the equilibrium volume and $c/a$ ratio
	of the cohesive energy $E^{\rm coh}$ at 0\,K.}
	\label{tab:Fel}
	\begin{tabular}{rrrrr}
	\hline
	&\multicolumn{4}{c}{Free energies (meV/site)}\\
	&\multicolumn{2}{c}{Pure Ti} & \multicolumn{2}{c}{Ti$_6$O}\\ 
	$T$(K) &  $F^{\rm el}$  & $F^{\rm vib}$  & $F^{\rm el}$ & $F^{\rm vib}$\\
	\hline
	   0 & $  0.0$ & $   0.7$ & $ 0.0$ & $  46.8$ \\
	 300 & $ -0.9$ & $ -45.7$ & $-1.5$ & $   5.9$ \\
	 600 & $ -4.1$ & $-169.2$ & $-3.8$ & $-115.8$ \\
	 900 & $-10.0$ & $-332.1$ & $-8.7$ & $-283.0$ \\
	\hline
	\end{tabular}
	\end{center}
\end{table}

The total free energy $F(V,\gamma,T) = E^{\rm coh}(V,\gamma) + F^{\rm vib}(V,\gamma,T)$
is obtained by adding the cohesive energy $E^{\rm coh}(V,\gamma)$ directly given by DFT calculations. 
One could also consider the contribution $F^{\rm el}$ of electronic excitations to this free energy 
\cite{Wolverton1995,Zhang2017,Cottura2018}, but calculations performed for pure Ti
and for the ordered Ti$_6$O compound show that this contribution can be neglected 
compared to the cohesive and vibrational contributions (Tab. \ref{tab:Fel}).
The same conclusion was reached by Argaman \etal{} in pure Ti \cite{Argaman2016}.

\begin{figure}[!bp]
	\centering
	a)\includegraphics[width=0.7\linewidth]{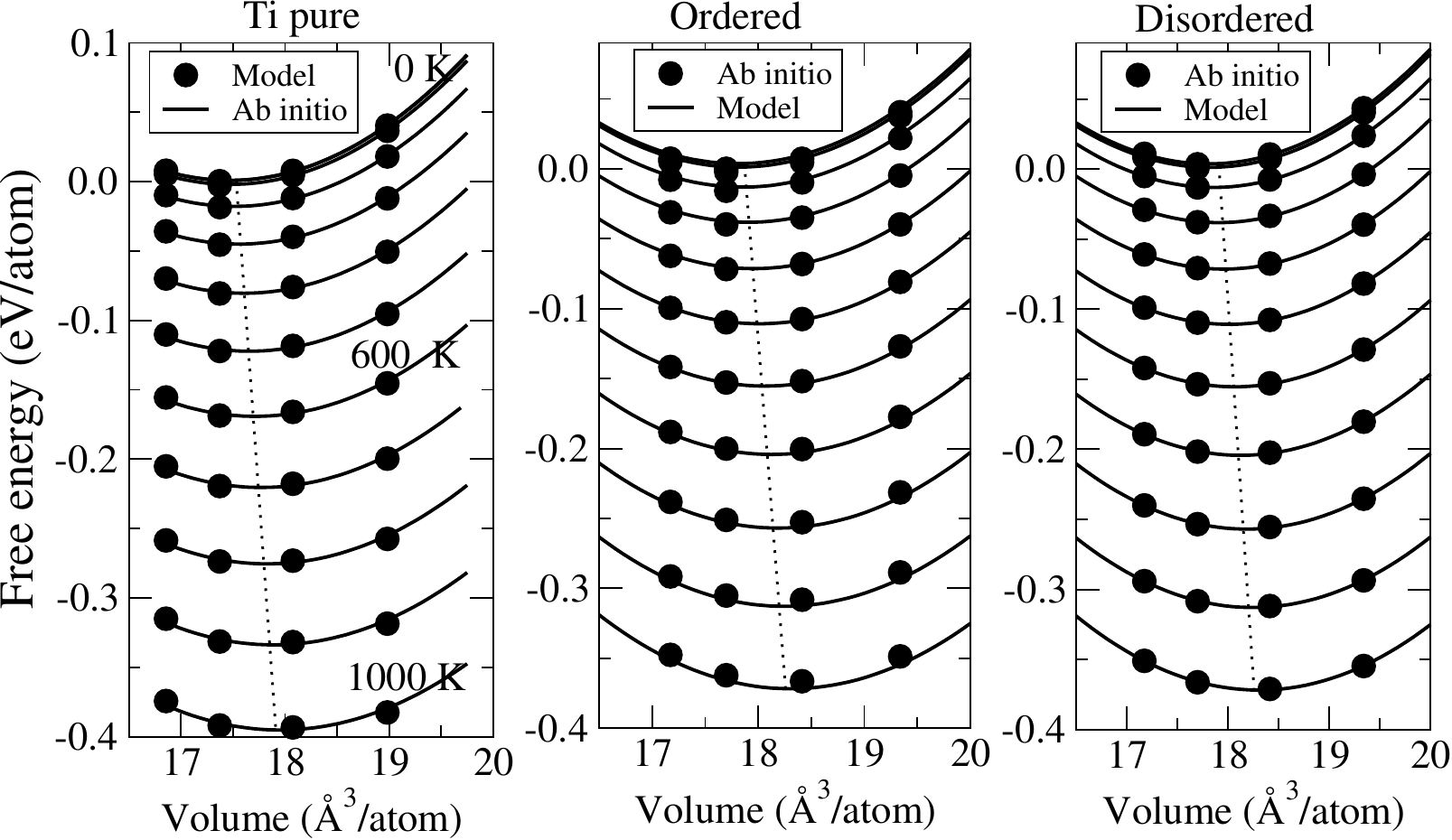}
	\\ \, \\ 
	b)\includegraphics[width=0.7\linewidth]{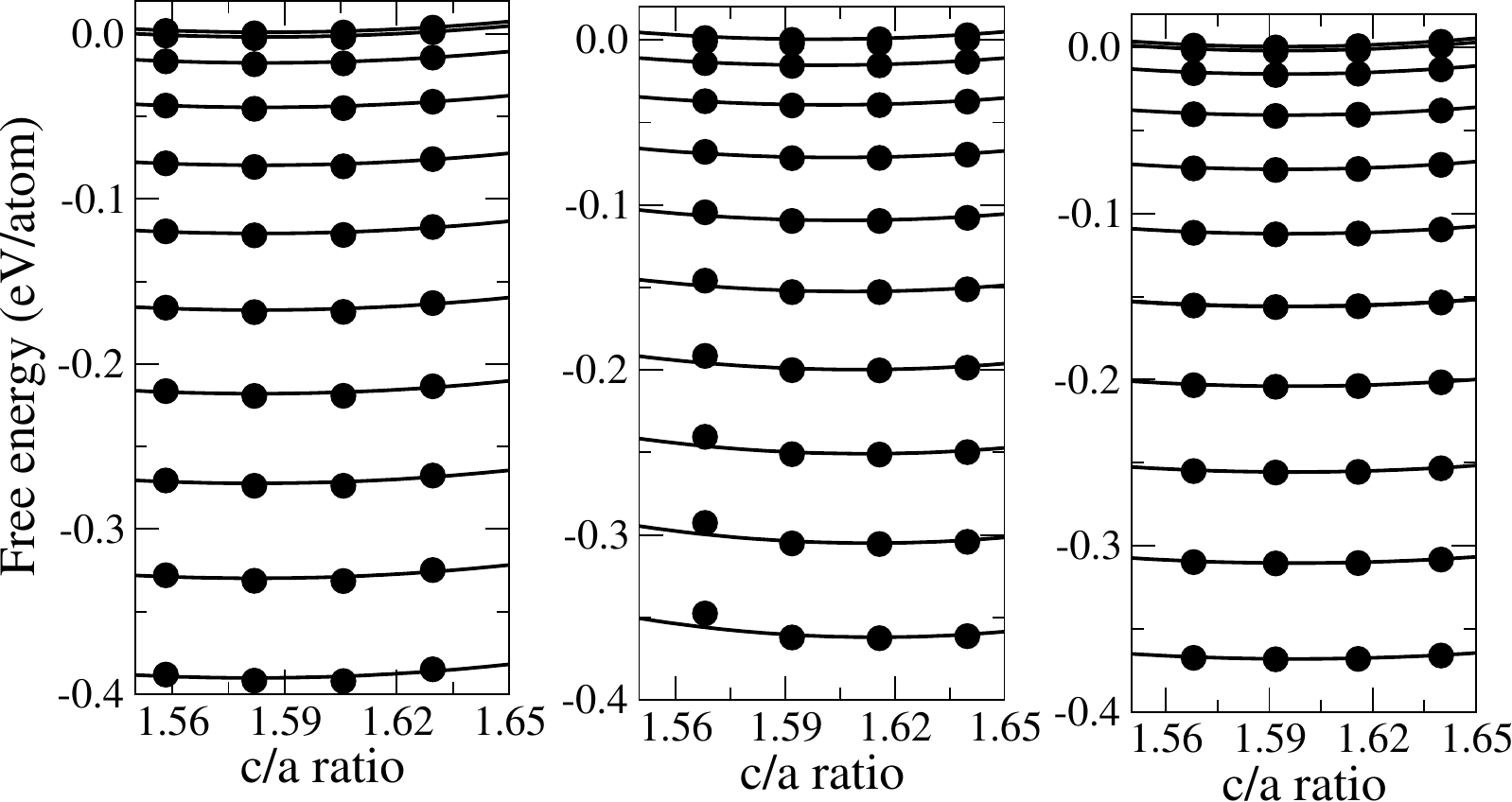}
	\caption{Variation of the free energy with a) the volume and b) the $\gamma=c/a$ ratio
	for pure Ti (left), the ordered Ti$_6$O compound (center) and a disordered solid solution of composition Ti$_6$O (right).
	Free energies are shown every 100\,K. 
	Symbols are the direct results of \abinitio{} calculations with quasi harmonic approximation 
	and lines their interpolation.
	The $c/a$ ratio and volume have been fixed to their 0\,K equilibrium values respectively in a) and b).
 	The dashed line in a) shows the position of the equilibrium volume for each temperature.}
	\label{fig:Fvib}
\end{figure}

For each structure, the phonon spectrum is calculated on a $4\times4$ grid corresponding to 4 different volumes $V$
and 4 different $\gamma=c/a$ ratios choosing a variation $\sim \pm10\%$ around the equilibrium values of the cohesive energy. 
The free energy can then be defined on this $(V,\gamma)$ grid for each temperature
and a continuous representation obtained thanks to a smooth interpolation with second rank polynomials 
(Fig. \ref{fig:Fvib}).
Minimization of this free energy finally leads to the equilibrium lattice parameters $a(T)$ and $c(T)$.
As shown in appendix \ref{sec:pureTi}, this approach predicts lattice expansion in pure Ti
in agreement with experimental data (Fig. \ref{fig:pureTi}).

\subsection{Variations of lattice parameters}

\begin{figure}[!p]
    \centering
	\includegraphics[width=0.8\linewidth]{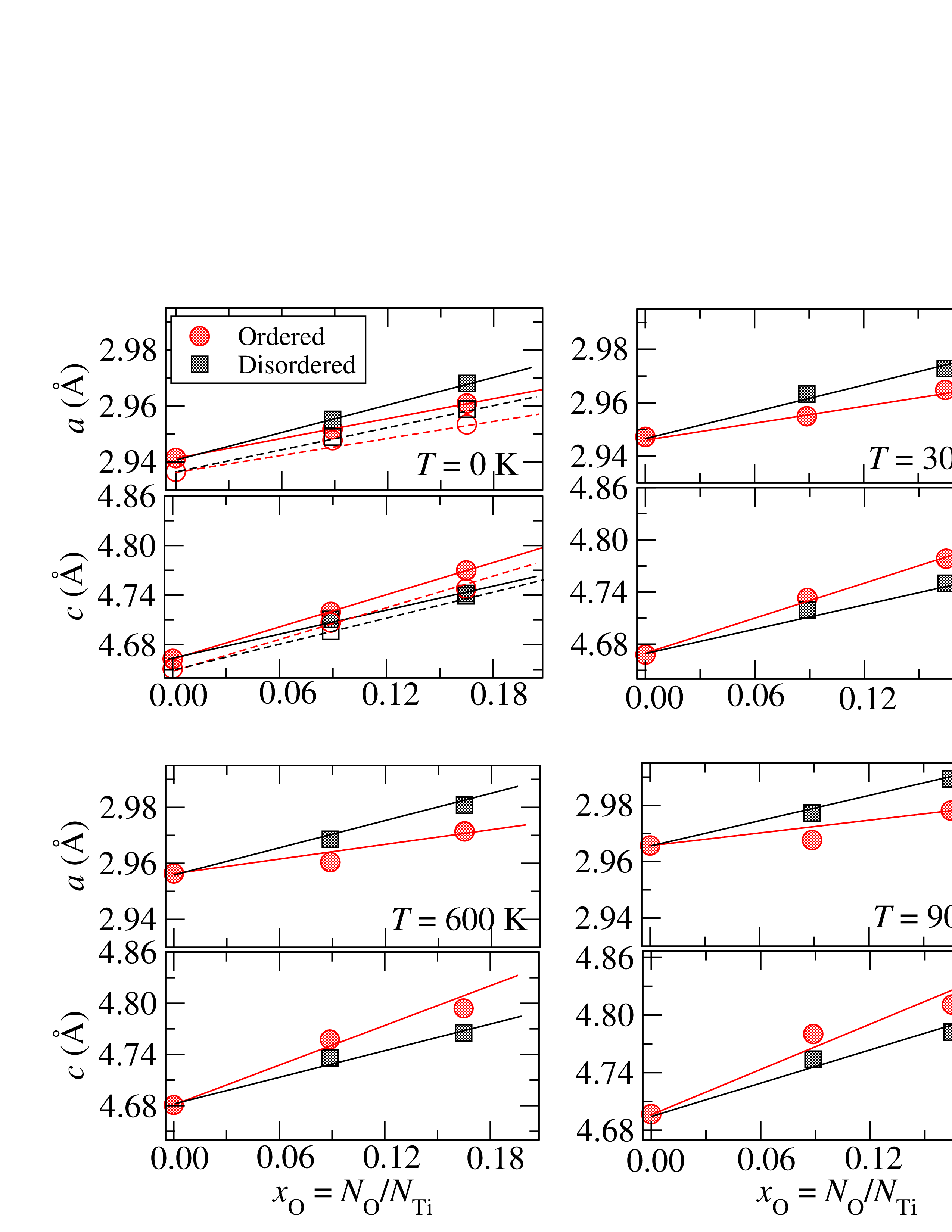}
	\caption{Lattice expansion in Ti-O alloys as a function of oxygen content $x_{\rm O}$. 
	Red disks correspond to ordered compounds and black squares to disordered solid solutions,
	both calculated for the compositions Ti$_{12}$O and Ti$_6$O. 
	Results in absence of atomic vibrations, without zero-point energy, are indicated at 0\,K 
	with open symbols and dashed lines.
	Solid and dashed lines are linear fits of the corresponding data, 
	imposing the lattice parameter of pure titanium in $x_{\rm O}=0$.
	}
	\label{fig:thermal_exp}
\end{figure}

Variations of lattice parameters with temperature are calculated 
for two different oxygen compositions, Ti$_6$O and Ti$_{12}$O, in addition to pure Ti. 
For each composition, we considered a random solid solution, as given by SQS method,
and a fully ordered structure with the most stable $AVBV$ stacking of basal planes
(Fig. \ref{fig:TinO_structures}).
At 0\,K, in absence of atomic vibrations, the same variations of the lattice parameters 
with O concentration and ordering state (open symbols in Fig. \ref{fig:thermal_exp}) are observed as in previous section
(Fig. \ref{fig:lattice_0K}), thus showing that the obtained lattice parameters do not depend on the supercell used
to model the different structures.
Accounting for atomic vibrations does not modify how these lattice parameters vary qualitatively.
Results confirm that lattice parameters increase with the oxygen content at all temperatures (Fig. \ref{fig:thermal_exp}), 
with variations which can be considered linear both in the disordered and the ordered states.
The impact of oxygen ordering is the same as the tendency observed at 0\,K for the highest concentrations:
the ordered state has larger $c$ and lower $a$ lattice parameter than the disordered solid solution
at the same oxygen concentration. This impact of oxygen ordering slightly increases with the temperature,
as can be seen from the values of the slopes $\alpha^{\rm O}_a=1/a\ \partial a/\partial x_{\rm O}$ and 
$\alpha^{\rm O}_c=1/c\ \partial c/\partial x_{\rm O}$ given in table \ref{tab:lattice_var_T}.
The difference between the values for the ordered and the disordered states increases with the temperature both for
$\alpha^{\rm O}_a$ and $\alpha^{\rm O}_c$.  But this variation
remains small and can be considered as second order compared to the variation with the oxygen concentration.

\begin{table}
	\centering
	\caption{Slopes characterizing linear variations of lattice parameters $a$ and $c$
	with O concentration, $\alpha^{\rm O}_a=1/a_{\rm Ti}\ \partial a/\partial x_{\rm O}$ and 
	$\alpha^{\rm O}_c=1/c_{\rm Ti}\ \partial c/\partial x_{\rm O}$, shown in Fig. \ref{fig:thermal_exp}
	for different temperatures.}
	\label{tab:lattice_var_T}
	\begin{tabular}{llllll}
		\hline
					&			& 0\,K	& 300\,K& 600\,K& 900\,K	\\
		\hline
		$\alpha^{\rm O}_a$	& Ordered TiO$_x$	& 0.040	& 0.035	& 0.028	& 0.022	\\
					& Disordered TiO$_x$	& 0.055	& 0.055	& 0.049	& 0.048	\\
		\hline
		$\alpha^{\rm O}_c$	& Ordered TiO$_x$	& 0.139	& 0.147	& 0.155	& 0.159	\\
					& Disordered TiO$_x$	& 0.106	& 0.111	& 0.115	& 0.116	\\
		\hline
	\end{tabular}
\end{table}

To a good approximation, one can neglect the variations with the temperature of the increase rates
$\alpha^{\rm O}_a$ and $\alpha^{\rm O}_c$.
The variations of the lattice parameters with the temperature and with the oxygen concentration 
are then reasonably well described by the equations
\begin{align}
	a(T,x_{\rm O}) &= a^0_{\rm Ti} \left( 1 + \alpha_a^T \, T \right)\left( 1 + \alpha_a^{\rm O} \, x_{\rm O} \right),
	\\
	c(T,x_{\rm O}) &= c^0_{\rm Ti} \left( 1 + \alpha_c^T \, T \right)\left( 1 + \alpha_c^{\rm O} \, x_{\rm O} \right),
	\label{eq:lattice_var}
\end{align}
showing that the increases with temperature $T$ and with oxygen concentration $x_{\rm O}$
can be factorized. 
Parameters given in Table \ref{tab:lattice_var} obtained through a fit of \abinitio{} calculations
confirm that $a$ and $c$ lattice parameters increase with the oxygen content, 
with a slightly higher (respectively lower) increase rate of the $c$ (respectively $a$) parameter
for the ordered than the disordered states.

\begin{table}
	\centering
	\caption{Parameters entering Eq. \ref{eq:lattice_var} and describing variations 
	of Ti lattice parameters with temperature and oxygen content
	for the ordered and disordered states.}
	\label{tab:lattice_var}
	\begin{tabular}{lll}
		\hline
		Pure Ti
		&$a^0_{\rm Ti} = 2.940$\,\AA			& $c^0_{\rm Ti} = 4.660$\,\AA\\
		&$\alpha_a^T = 9.4\times10^{-6}$\,K$^{-1}$	& $\alpha_c^T = 8.3\times10^{-6}$\,K$^{-1}$ \\
		Ordered TiO$_x$
		&$\alpha_a^{\rm O}=0.031$			& $\alpha_c^{\rm O} = 0.150$ \\
		Disordered TiO$_x$
		&$\alpha_a^{\rm O}=0.052$			& $\alpha_c^{\rm O} = 0.112$ \\
		\hline
	\end{tabular}
\end{table}

\clearpage
\section{Experiments and discussion}

\subsection{Material and methods}

X-ray diffraction (XRD) measurements are performed on titanium 
and a Ti-O binary alloy. 
The titanium, hereafter denoted pure titanium, contains only 600\,wppm oxygen, 
corresponding to an atomic composition $c_{\rm O}=N_{\rm O}/(N_{\rm Ti}+N_{\rm O})=0.18$\,at.\%
and to an occupation of interstitial octahedral sites $x_{\rm O}=N_{\rm O}/N_{\rm Ti}=0.18$\%,
while the oxygen nominal concentration in the Ti-O alloy is 6000\,wppm 
($c_{\rm O}=1.77$\,at.\% and $x_{\rm O}=1.81$\%).
These concentrations are confirmed by inert gas fusion analysis.
Both pure Ti and Ti-O alloy are first hot- and cold-rolled with a thickness reduction 
of 75\% and 40\% respectively to finally obtain 1\,mm thick plates. 
The samples are then recrystallized in molten salt baths 
(Li$_2$CO$_3$, Na$_2$CO$_3$, K$_2$CO$_3$) at 1023\,K during 600\,s. 
After water quenching, both pure titanium and Ti-O samples present fully recrystallized $\alpha$ grains
with a 90 and 30\,$\mu$m size respectively. 
The samples are then polished with SiC grinding papers, 
before a final polishing with colloidal silica suspension.

XRD patterns are obtained with an XPert Pro Panalytical instrument operating at 40\,kV and 40\,mA. 
A cobalt source ($K_{\alpha1}=1.789$\,\AA{} and $K_{\alpha2}=1.793$\,\AA) is used. A first diffractogram is recorded at room temperature
without the heating chamber, so that the sample height can be adjusted 
after installing the chamber by comparing the peak positions. 
After the measurement at room temperature, an Anton Paar HTK 1200N furnace with a resistive system is used to heat the sample holder
up to 473\,K 
and then to different temperatures between 473 and 773\,K at 10\,K/min.
The temperature measured on the sample holder could slightly differ from the sample temperature.
The heating chamber is kept under argon flow to prevent oxidation.
After one hour of heating stabilization for each temperature,
diffractograms are recorded in the 40-90\degree{} $2\theta$-range with a 0.016\degree{} step size
during 80\,s.  5 scans are recorded at each temperature step and then summed up for the analysis.

Lattice parameters of both $\alpha$ matrix and ordered precipitates are determined 
thanks to a Rietveld refinement using the \maud{} software.
The refinement takes account of the non-monochromatic feature of the X-ray beam and allows to assign the shoulder on the right of the peaks 
to the $K_{\alpha2}$ ray of the cobalt source.
The Caglioti parameters of the diffractometer are first calibrated with a LaB$_6$ powder.
The COD (Crystallographic Open Database) ID \#9008517 is chosen for the crystal structure of the $\alpha$ matrix. 
Ordered precipitates have a Ti$_6$O structure, which corresponds to an ordering of the oxygen atoms 
in the octahedral interstitial sites of the hcp Ti-lattice \cite{Poulain2022}.
XRD pattern of the precipitate phase should theoretically display the same principal peaks as the $\alpha$ matrix,
with some additional peaks arising from oxygen ordering. 
However, a previous study on the same Ti-O samples \cite{Poulain2022} has shown that these additional peaks cannot be detected by XRD
because of their too low intensity
and that Ti$_6$O-type precipitates can be characterized only by shoulders on the left of the matrix peaks,
corresponding to slightly larger $a$ and $c$ lattice parameters of the underlying hcp lattice. 
Therefore, the same COD ID is used in the Rietveld refinement for the ordered precipitates as for the $\alpha$ matrix.
Finally, the sample holder in alumina leads to some additional peaks which are taken into account in the Rietveld refinement
by considering an Al$_2$O$_3$ phase (COD ID \#1000059).

Transmission electron microscopy (TEM) observations of the Ti-O alloys 
are performed with a JEOL TEM after the recrystallization heat treatment,
using the same sample preparation as in Refs. \cite{Poulain2022,Amann2023}.

\subsection{Results}

\begin{figure}[!bp]
	\centering
	\includegraphics[width=0.7\linewidth]{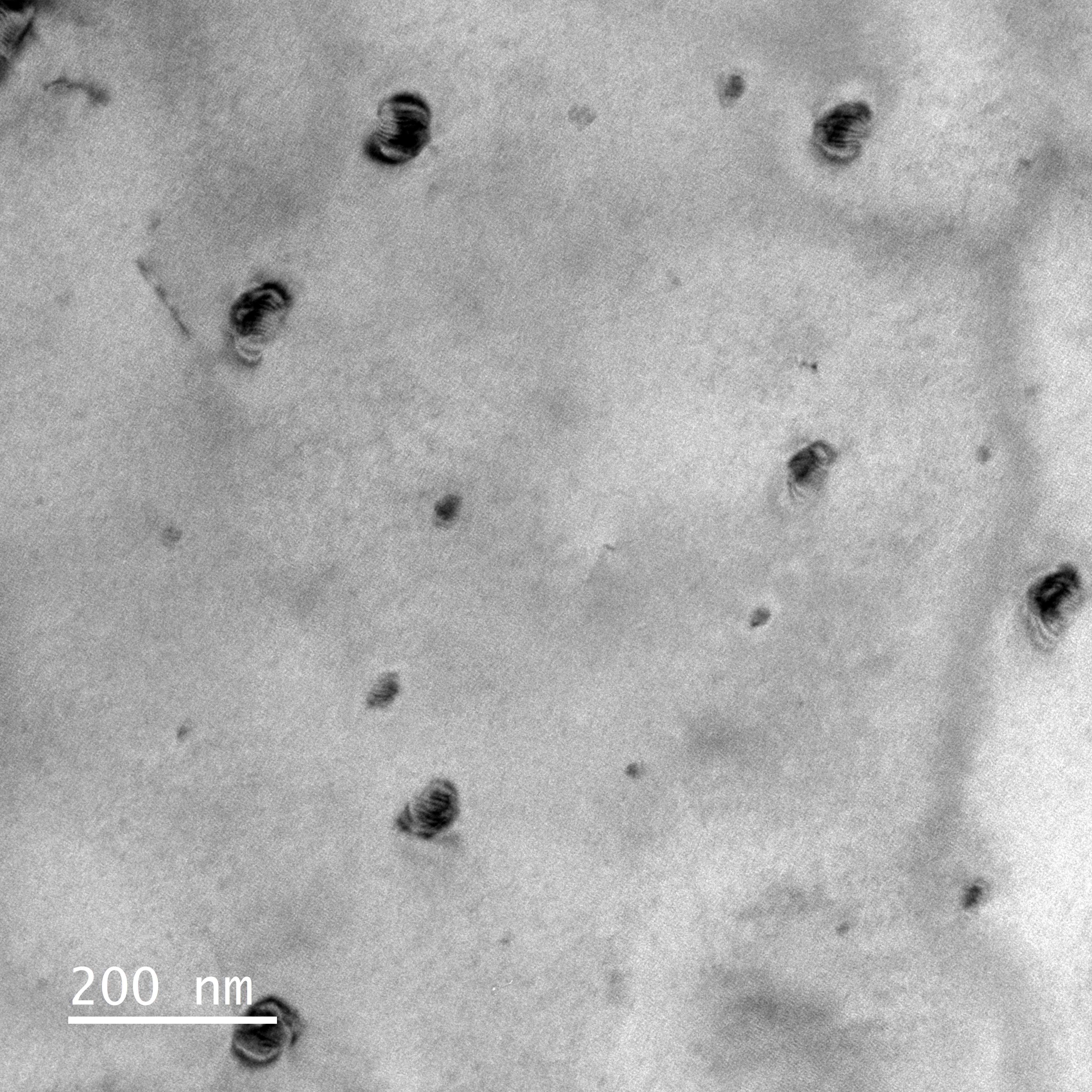}
	\caption{TEM observation in bright field mode of the Ti-O alloy.}
	\label{fig:TEM}
\end{figure}

TEM observations confirm the presence of precipitates with a nanometric size (Fig. \ref{fig:TEM}).
These precipitates, which have been previously imaged in dark-field conditions using 
the superlattice diffraction of the Ti$_6$O ordered structure \cite{Poulain2020,Poulain2022},
can also be imaged in bright field mode, thanks to their elastic strain field.
This strain field arises from the lattice misfit between the ordered precipitates and the surrounding titanium matrix.

\begin{figure}[!bp]
	\centering
	\includegraphics[width=0.6\linewidth]{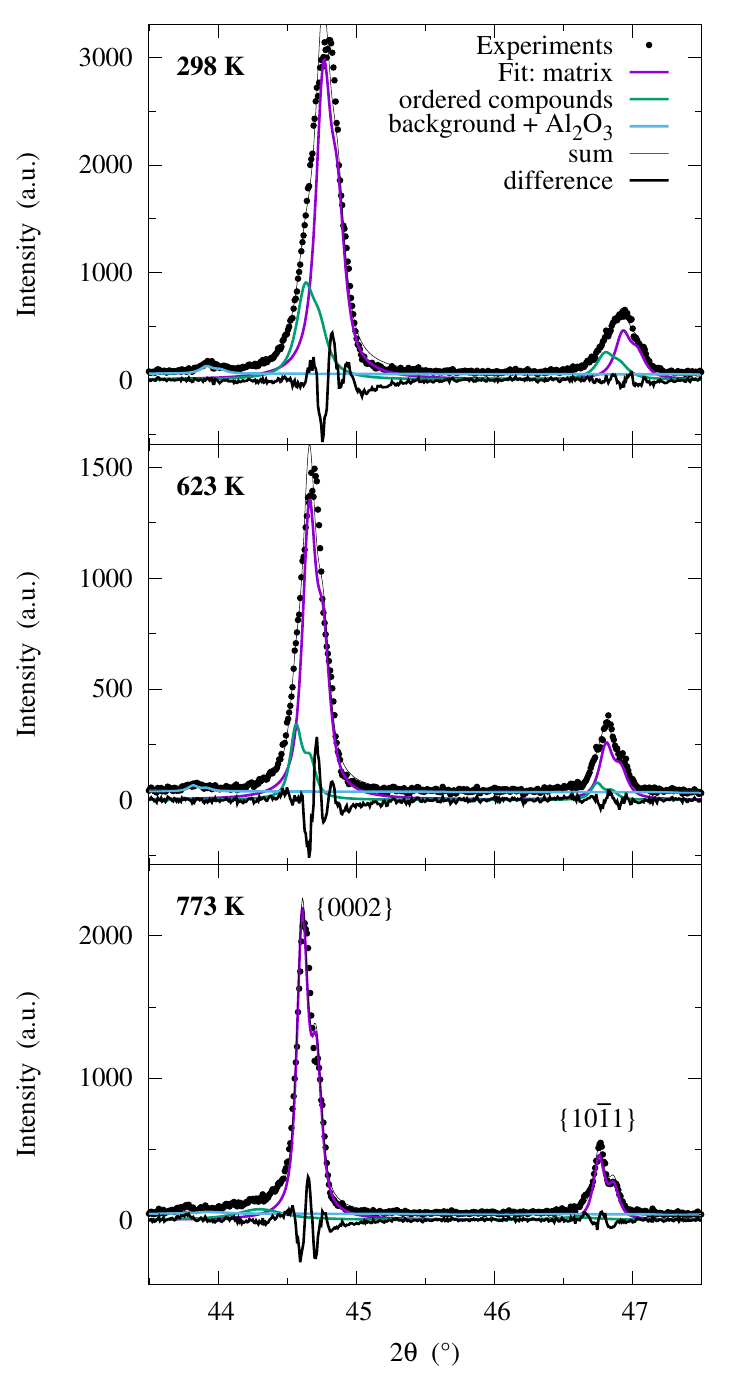}
	\caption{X-ray diffraction patterns obtained at three different temperatures
	and their Rietveld refinement showing the different contributions of the $\alpha$-Ti matrix,
	the ordered compounds, and the alumina sample holder.
	These refinements lead to goodness of fit (GOF) values \cite{Toby2006} between 1.1 and 1.5.}
	\label{fig:XRD}
\end{figure}

This lattice misfit leads to the presence of shoulders on the diffraction peaks in the XRD diffractograms 
(Fig. \ref{fig:XRD} and Fig. \ref{fig:XRD_all} in appendix \ref{sec:XRD}).
This shoulder can be clearly seen on the left of the diffraction peak
corresponding to the basal \hkl{0002} planes ($44^{\circ} < 2\theta < 45^{\circ}$).
It is the signature of the larger $c$ lattice parameter of the ordered compounds
compared to the one of the matrix. 
Note that this shoulder appearing for the basal  \hkl{0002} planes cannot be explained 
by a martensitic transformation of the hexagonal $\alpha$ phase to the orthorhombic $\alpha''$ phase 
as such a transformation will lead to a splitting of all diffraction peaks except the ones corresponding to basal planes.
This shoulder on \hkl{0002} diffraction peaks can be explained only by a mixture of two phases 
with different $c$ lattice parameters.
Such a left shoulder is also visible on the peak corresponding 
to pyramidal \hkl{10-11} planes ($46.5^{\circ} < 2\theta < 47^{\circ}$). 
On the other hand, because of the sample texture, the intensity of the peaks corresponding 
to \hkl{10-10} prismatic planes is not high enough to detect any shoulder. 
It is also observed that  the signature of the ordered compounds is not very intense 
and almost disappears at 723\,K and above (Fig. \ref{fig:XRD_all} in appendix \ref{sec:XRD}),
making analysis at these temperatures difficult.  
This is an indication that the ordered compounds have started to dissolve
and that O solubility limit at these temperatures is higher than the sample nominal concentration,
\ie{} 6000\,wppm, which looks reasonable according to the theoretical phase diagram \cite{Gunda2018}.
Such a dissolution of ordered precipitates have also been seen recently in a Ti-Zr-O alloy using TEM and resistivity
with \insitu{} heating experiments \cite{Amann2023p}.

\begin{figure}[!bt]
	\centering
	\includegraphics[width=0.49\linewidth]{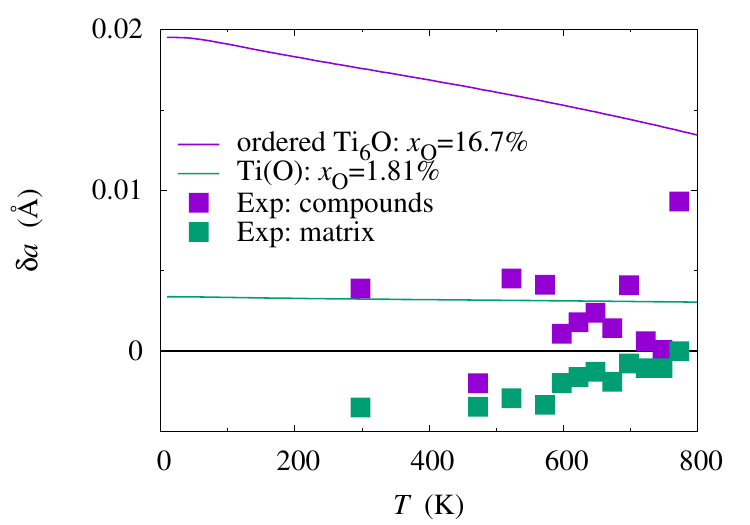}
	\hfill
	\includegraphics[width=0.49\linewidth]{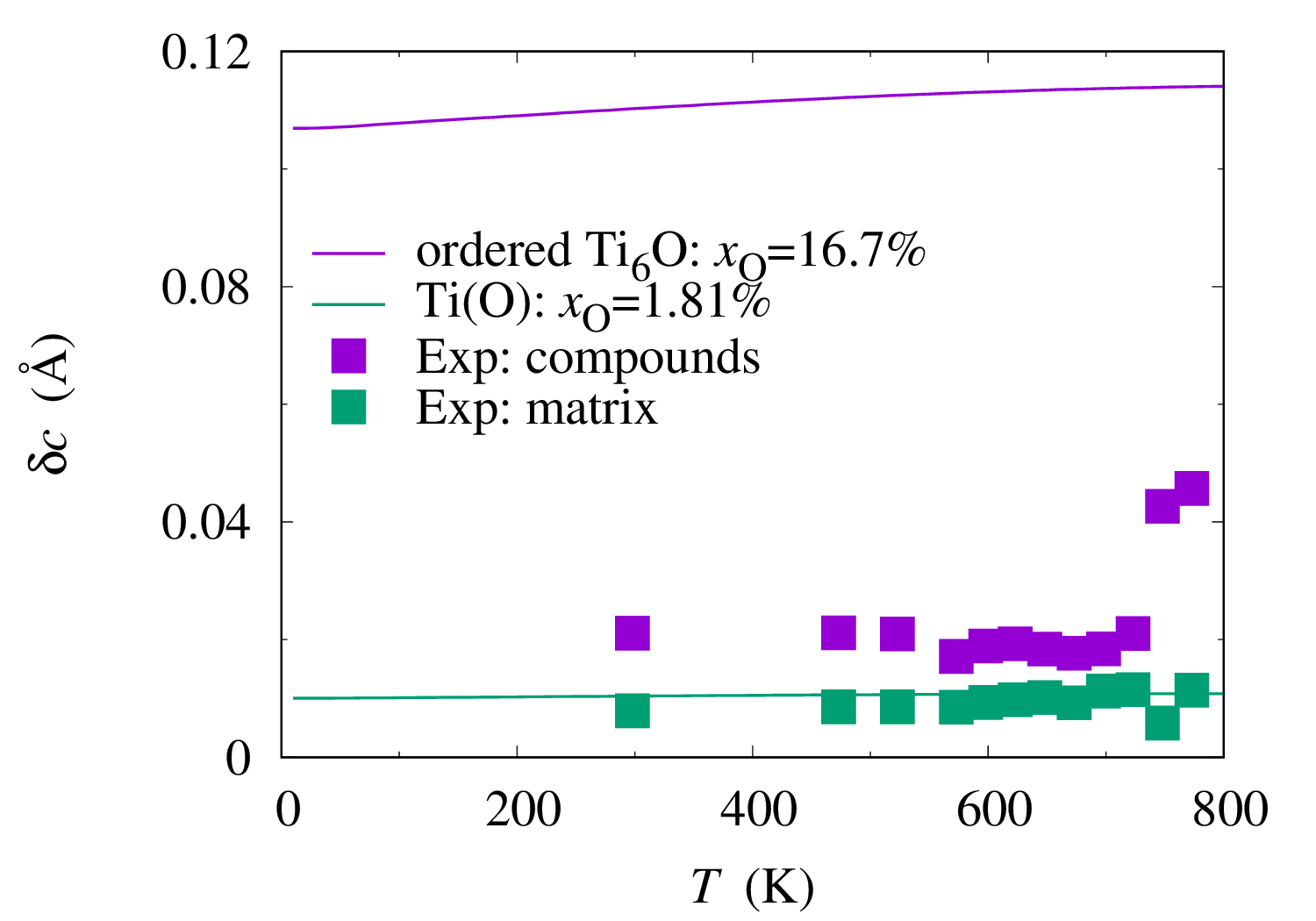}
	\caption{Increase of lattice parameters $a$ and $c$ induced by oxygen as a function of the temperature.
	Lines are the predictions of the \abinitio{} model for the ordered Ti$_6$O ordered compound (purple line)
	and for a random solid solution of composition $\xO=1.81$\% (green line).
	Symbols are results of experiments for the ordered compounds and for the matrix.}
	\label{fig:comparison_exp}
\end{figure}

\begingroup
\squeezetable
\begin{table}[!pb]
	\caption{Experimental values of lattice parameters deduced from \maud{} Rietveld refinement 
	and oxygen content $\xO = N_{\rm O}/N_{\rm Ti}$ deduced from these measurements using Eq. \ref{eq:xO_exp}.}
	\label{tab:experiment}
	\centering
	\begin{tabular}{cccrrrrrrc}
		\hline
		$T$ (K)	&sample		&  phase	& \multicolumn{3}{c}{$a$ (\AA{})}	& \multicolumn{3}{c}{$c$ (\AA{})} 	& $\xO$ \\
		\hline
298	& pure Ti	& matrix	& \ 2.9536	& $\pm$ & $ 1\times10^{-4}$		& \ 4.6850    	& $\pm$ & $1\times10^{-4}$	\\
	& Ti-O		& matrix	& \ 2.9501	& $\pm$ & $ 1\times10^{-4}$		& \ 4.6929  	& $\pm$ & $1\times10^{-4}$	& \ 0.012 \\
	& Ti-O		& compounds	& \ 2.9575	& $\pm$ & $ 2\times10^{-4}$		& \ 4.7061  	& $\pm$ & $2\times10^{-4}$	& \ 0.030   \\
473	& pure Ti	& matrix	& \ 2.9586	& $\pm$ & $ 1\times10^{-4}$		& \ 4.6913	& $\pm$ & $1\times10^{-4}$	\\  
	& Ti-O		& matrix	& \ 2.9551	& $\pm$ & $ 1\times10^{-4}$		& \ 4.6999	& $\pm$ & $1\times10^{-4}$	& \ 0.013 \\
	& Ti-O		& compounds	& \ 2.9566	& $\pm$ & $ 3\times10^{-4}$		& \ 4.7125	& $\pm$ & $2\times10^{-4}$	& \ 0.029 \\
523	& pure Ti	& matrix	& \ 2.9596	& $\pm$ & $ 1\times10^{-4}$		& \ 4.6937	& $\pm$ & $1\times10^{-4}$	\\  
	& Ti-O		& matrix	& \ 2.9567	& $\pm$ & $ 1\times10^{-4}$		& \ 4.7022	& $\pm$ & $1\times10^{-4}$	& \ 0.013 \\
	& Ti-O		& compounds	& \ 2.9642	& $\pm$ & $ 6\times10^{-4}$		& \ 4.7146	& $\pm$ & $2\times10^{-4}$	& \ 0.030 \\
573	& pure Ti	& matrix	& \ 2.9610	& $\pm$ & $ 1\times10^{-4}$		& \ 4.6960	& $\pm$ & $1\times10^{-4}$	\\  
	& Ti-O		& matrix	& \ 2.9577	& $\pm$ & $ 1\times10^{-4}$		& \ 4.7045	& $\pm$ & $1\times10^{-4}$	& \ 0.013 \\
	& Ti-O		& compounds	& \ 2.9651	& $\pm$ & $ 7\times10^{-4}$		& \ 4.7132	& $\pm$ & $2\times10^{-4}$	& \ 0.025 \\
598	& pure Ti	& matrix	& \ 2.9613	& $\pm$ & $ 1\times10^{-4}$		& \ 4.6966	& $\pm$ & $1\times10^{-4}$	\\  
	& Ti-O		& matrix	& \ 2.9594	& $\pm$ & $ 1\times10^{-4}$		& \ 4.7060 	& $\pm$ & $1\times10^{-4}$	& \ 0.015 \\
	& Ti-O		& compounds	& \ 2.9624	& $\pm$ & $ 4\times10^{-4}$		& \ 4.7156	& $\pm$ & $2\times10^{-4}$	& \ 0.027 \\
623	& pure Ti	& matrix	& \ 2.9616	& $\pm$ & $ 1\times10^{-4}$		& \ 4.6977   	& $\pm$ & $1\times10^{-4}$	\\  
	& Ti-O		& matrix	& \ 2.9600	& $\pm$ & $ 1\times10^{-4}$		& \ 4.7075  	& $\pm$ & $1\times10^{-4}$	& \ 0.016  \\
	& Ti-O		& compounds	& \ 2.9634	& $\pm$ & $ 3\times10^{-4}$		& \ 4.7170  	& $\pm$ & $2\times10^{-4}$	& \ 0.027  \\
648	& pure Ti	& matrix	& \ 2.9621	& $\pm$ & $ 1\times10^{-4}$		& \ 4.6986 	& $\pm$ & $1\times10^{-4}$	\\  
	& Ti-O		& matrix	& \ 2.9608	& $\pm$ & $ 1\times10^{-4}$		& \ 4.7087	& $\pm$ & $1\times10^{-4}$	& \ 0.017 \\
	& Ti-O		& compounds	& \ 2.9645	& $\pm$ & $ 3\times10^{-4}$		& \ 4.7169 	& $\pm$ & $2\times10^{-4}$	& \ 0.026 \\
673	& pure Ti	& matrix	& \ 2.9626	& $\pm$ & $ 1\times10^{-4}$		& \ 4.6993 	& $\pm$ & $1\times10^{-4}$	\\  
	& Ti-O		& matrix	& \ 2.9607	& $\pm$ & $ 1\times10^{-4}$		& \ 4.7086	& $\pm$ & $1\times10^{-4}$	& \ 0.015 \\
	& Ti-O		& compounds	& \ 2.9640	& $\pm$ & $ 3\times10^{-4}$		& \ 4.7170	& $\pm$ & $2\times10^{-4}$	& \ 0.025 \\
698	& pure Ti	& matrix	& \ 2.9630	& $\pm$ & $ 1\times10^{-4}$		& \ 4.7002 	& $\pm$ & $1\times10^{-4}$	\\  
	& Ti-O		& matrix	& \ 2.9623	& $\pm$ & $ 1\times10^{-4}$		& \ 4.7115	& $\pm$ & $1\times10^{-4}$	& \ 0.019 \\
	& Ti-O		& compounds	& \ 2.9671	& $\pm$ & $ 5\times10^{-4}$		& \ 4.7186	& $\pm$ & $4\times10^{-4}$	& \ 0.026 \\
723	& pure Ti	& matrix	& \ 2.9636	& $\pm$ & $ 1\times10^{-4}$		& \ 4.7014 	& $\pm$ & $1\times10^{-4}$	\\  
	& Ti-O		& matrix	& \ 2.9625	& $\pm$ & $ 1\times10^{-4}$		& \ 4.7128	& $\pm$ & $1\times10^{-4}$	& \ 0.019 \\
	& Ti-O		& compounds	& \ 2.9642	& $\pm$ & $ 1\times10^{-4}$		& \ 4.7224	& $\pm$ & $4\times10^{-4}$	& \ 0.029 \\
748	& pure Ti	& matrix	& \ 2.9642	& $\pm$ & $ 1\times10^{-4}$		& \ 4.7025	& $\pm$ & $1\times10^{-4}$	\\  
	& Ti-O		& matrix	& \ 2.9632	& $\pm$ & $ 1\times10^{-4}$		& \ 4.7083	& $\pm$ & $1\times10^{-4}$	& \ 0.010 \\
	& Ti-O		& compounds	& \ 2.9643	& $\pm$ & $ 1\times10^{-4}$		& \ 4.7451	& $\pm$ & $7\times10^{-4}$	& \ 0.059  \\
773	& pure Ti	& matrix	& \ 2.9646	& $\pm$ & $ 1\times10^{-4}$  		& \ 4.7036   	& $\pm$ & $1\times10^{-4}$	\\  
	& Ti-O		& matrix	& \ 2.9646  	& $\pm$ & $ 1\times10^{-4}$	 	& \ 4.7150  	& $\pm$ & $1\times10^{-4}$	& \ 0.020 \\
	& Ti-O		& compounds	& \ 2.9739  	& $\pm$ & $ 4\times10^{-4}$ 		& \ 4.7493  	& $\pm$ & $6\times10^{-4}$	& \ 0.065   \\		
	\hline
	\end{tabular}
\end{table}
\endgroup

Rietveld refinement of XRD diffractograms allows extracting the lattice parameters $a$ and $c$ 
of pure Ti, as well as of both the matrix and the ordered compounds in the Ti-O alloy (Tab. \ref{tab:experiment}).
As expected, the ordered compounds have larger lattice parameters than the matrix in the Ti-O alloy 
for all the considered temperatures. 
To compare the lattice parameters measured in the Ti-O alloy with the ones deduced from \abinitio{} calculations, 
we compute the differences $\delta a$ and $\delta c$ between the parameters measured in the alloy 
and in pure Ti at the same temperature, as these variations are less sensitive to the choice 
of the exchange-correlation functional used for \abinitio{} calculations than absolute values $a$ and $c$
(\cf{} appendix \ref{sec:LDA}). 
We neglect, for experimental data, the presence of 0.18\,at.\% oxygen in pure Ti, 
as it only has a marginal impact on the obtained $\delta a$ and $\delta c$ values.
$\delta a$ and $\delta c$ characterize the increase of lattice parameters
induced by the presence of oxygen.  
The increase $\delta c$ measured for the matrix phase is almost constant with the temperature 
and its value corresponds to the theoretical one obtained for a random solid solution with the same nominal 
concentration as the Ti-O alloy (Fig. \ref{fig:comparison_exp}). 
This is a strong indication that there is no real oxygen depletion in the matrix 
despite the presence of ordered compounds.  
The experimental lattice increases $\delta c$ obtained for the compounds are higher than the ones of the matrix
but do not reach the theoretical values expected for ordered compounds with a Ti$_6$O stoichiometry (Fig. \ref{fig:comparison_exp}). 
Although these compounds are enriched in oxygen, they do not have the expected 1/6 stoichiometry.
The same conclusions could be drawn from the obtained $\delta a$ lattice increase although things are less clear. 
The uncertainty comes from the smallest expected values for $\delta a$ than for $\delta c$, 
because the impact of oxygen is smaller on $a$ than $c$ as already shown before (Fig. \ref{fig:thermal_exp}).  
Besides, because of the sample texture leading to the absence of the prismatic peak, 
the experimental precision on $a$ is not as high as on $c$, 
explaining why slightly negative values are obtained for $\delta a$ in the matrix
while a lattice increase is expected.

\subsection{Discussion}

\begin{figure}[!bt]
	\centering
	\includegraphics[width=0.6\linewidth]{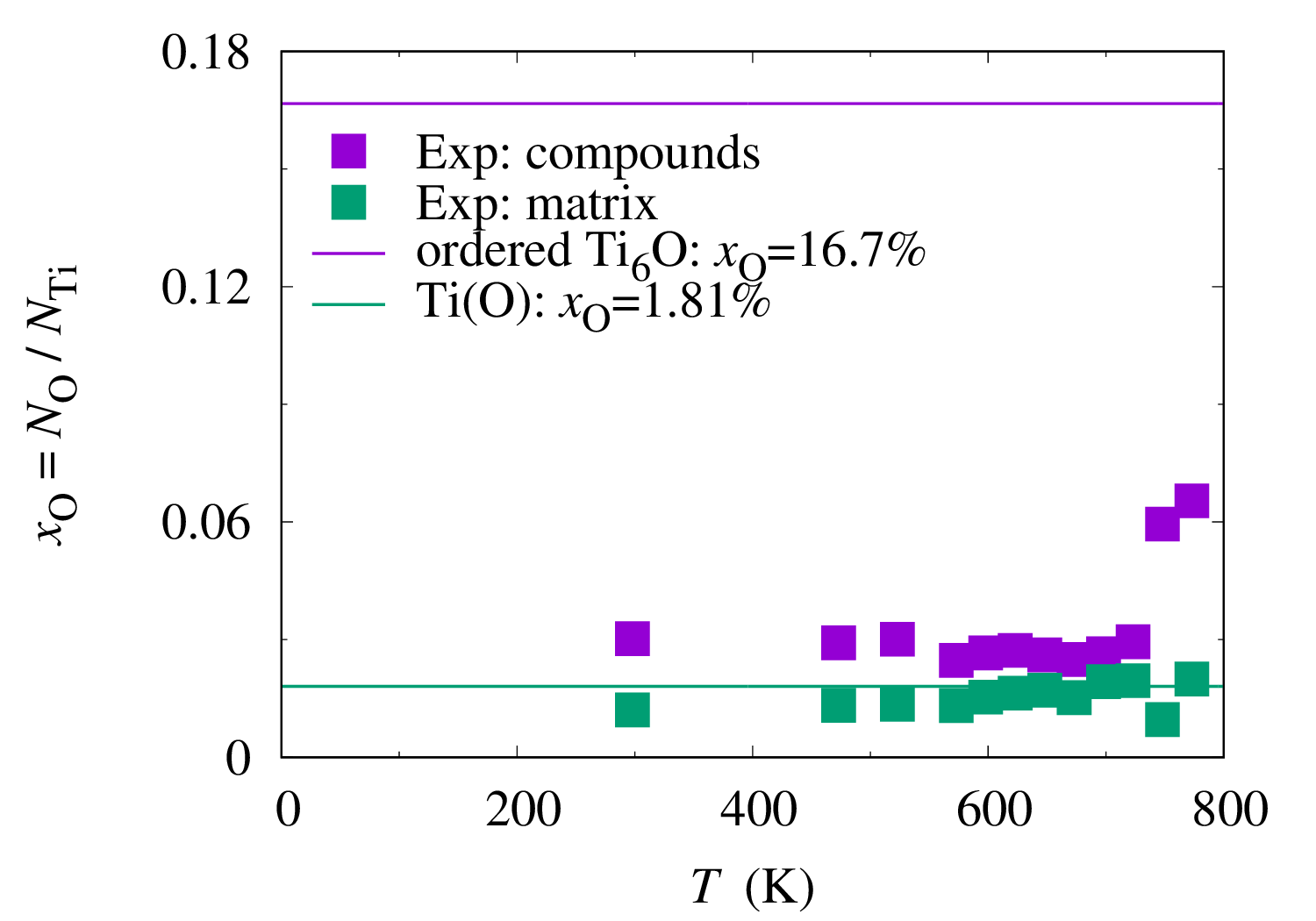}
	\caption{Oxygen content $\xO$ of the matrix and of the ordered compounds
	determined with Eq. \ref{eq:xO_exp} from the experimental lattice parameters (Tab. \ref{tab:experiment})
	and from the theoretical increase rates $\alpha_{a}^{\rm O}$ and $\alpha_{c}^{\rm O}$ (Tab. \ref{tab:lattice_var}).
	The horizontal lines correspond to the nominal composition $\xO=0.0181$ of the Ti-O alloy
	and to the composition of stoichiometric Ti$_6$O compounds ($\xO=1/6\sim0.167$).}
	\label{fig:xO_exp}
\end{figure}

Despite the uncertainty existing on both the theoretical and experimental variations
of Ti lattice parameters with oxygen content and with ordering state, one can try to determine the composition 
of the ordered compounds and of the matrix from the increases $\delta a$ and $\delta c$ 
of the lattice parameters, which have been measured by XRD at the different temperatures. 
A least square fit of both $a$ and $c$ parameters leads to the following expression for the occupation 
of the octahedral interstitial sites
\begin{equation}
	\xO(\delta a,\delta c) = \frac{ a_{\rm Ti} \, \alpha_{a}^{\rm O} \, \delta a + c_{\rm Ti} \, \alpha_{c}^{\rm O} \, \delta c }
	{ \left( a_{\rm Ti} \, \alpha_{a}^{\rm O} \right)^2 + \left( c_{\rm Ti} \, \alpha_{c}^{\rm O} \right)^2 }.
	\label{eq:xO_exp}
\end{equation}
Using the lattice parameters of pure Ti, $a_{\rm Ti}$ and $c_{\rm Ti}$, determined experimentally at each temperature
with the theoretical lattice increase $\alpha_{a}^{\rm O}$ and $\alpha_{c}^{\rm O}$ given in table \ref{tab:lattice_var}, 
thus neglecting their variations with the temperature,
one obtains the compositions $\xO$ given in table \ref{tab:experiment} and shown in Fig. \ref{fig:xO_exp}.
These results confirm that the ordered compounds have a lower composition than the expected value $\xO=1/6\sim0.167$
corresponding to Ti$_6$O stoichiometry. 
Their composition starts to rise at the two highest temperatures considered in this study, 748 and 773\,K, 
but without reaching the Ti$_6$O composition. 
However, as noted before, the intensity of the diffraction peaks assigned to the ordered compounds is not very intense
at these two temperatures (Fig. \ref{fig:XRD_all} in appendix \ref{sec:XRD}) and one cannot expect an accurate determination 
of the corresponding lattice parameters at these high temperatures where the ordered compounds have probably started to dissolve.
Fig. \ref{fig:xO_exp} also illustrates that this approach to determine the composition of the matrix and of the ordered compounds
cannot be considered fully quantitative: at low temperature, one observes an oxygen depletion of the matrix and an enrichment of the compounds, as expected, 
but at the highest temperatures, both phases are enriched, which violates conservation law. 
This uncertainty on the obtained composition arises from the difficulty to deconvoluate the X-ray diffraction pattern
to associate different lattice parameters to different phases, the closer the lattice parameters, the more difficult. 
It is also hard to reach a precision high enough in the \abinitio{} determination of the increase rates 
$\alpha_{a}^{\rm O}$ and $\alpha_{c}^{\rm O}$, especially for such small variations of the lattice parameters. 
Finally, when deriving a composition from measured lattice parameters, we assume that the different phases are free of stress, 
which is not true for coherent compounds embedded in a matrix.
Nevertheless, despite all the remaining uncertainties, the approach allows concluding that the ordered compounds do not have 
the Ti$_6$O composition and are only slightly enriched with oxygen compared to the surrounding matrix. 

As shown in Ref. \cite{Poulain2022} with electronic diffraction, these ordered compounds, which can be found in binary Ti-O alloys,
have a crystallographic structure corresponding to the one known for Ti$_6$O crystal (space group P31c). 
Although the binary phase diagram predicts that the equilibrium composition of these ordered compounds should be close to 
Ti$_6$O stoichiometry \cite{Gunda2018}, the present work shows that these precipitates do not have this composition. 
One possibility is that oxygen diffusion is not fast enough to reach thermal equilibrium.
Considering the oxygen diffusion coefficient $D_{\rm O}(T) = D^0_{\rm O} \exp{(-Q_{\rm O}/R\,T)}$ 
with an activation energy  $Q_{\rm O}=200$\,kJ\,mol$^{-1}$
and a diffusion pre-exponential constant $D^0_{\rm O} = 4.5\times10^{-5}$\,m$^2$\,s$^{-1}$ \cite{Liu1988},
one can estimate the mean square displacement of oxygen atoms when diffusing a time laps $\Delta t$,
$d_{\rm O}(T,\Delta t) = \sqrt{6\,D_{\rm O}(T)\,\Delta t}$.
For the recrystallization heat treatment ($T=1023$\,K and $\Delta t=600$\,s), 
the oxygen mean square displacement is 3\,$\mu$m, thus showing that equilibrium is reached. 
But, according to the theoretical phase diagram \cite{Gunda2018}, this temperature is too high for Ti$_6$O ordered compounds 
to be stable.  The disappearance of the signature of this ordered phase in the diffraction patterns above 723\,K  
is also a strong indication that Ti$_6$O compounds are not stable at the recrystallization temperature.
The ordered compounds should rather appear during the quench, when the temperature becomes 
low enough to lead to the decomposition of the initial homogeneous solid solution 
in a depleted solid solution and an ordered phase with Ti$_6$O-type crystallographic structure. 
If the subsequent quench is fast enough, which should be true for the water quench used here,
one does not expect that the forming precipitates reach equilibrium, thus explaining why the observed ordered compounds 
have a lower composition than the expected Ti$_6$O stoichiometry. 

Considering now the stays ($\Delta t=3600$\,s) at different temperatures for the XRD measurements, 
the composition of the precipitates remains almost unchanged up to 723\,K (Fig. \ref{fig:xO_exp}), 
with the signature of the ordered compounds in the XRD pattern disappearing at this temperature
and above (Fig. \ref{fig:XRD_all} in appendix \ref{sec:XRD}).
Oxygen diffusion should be too slow below 723\,K to allow for a change of precipitates.
This appears reasonable as the oxygen mean square displacement is only 60\,nm at $T=723$\,K.
At this temperature and above, it appears high enough 
to allow for precipitate evolution, in particular their dissolution, but not below.
Better characterization of the kinetic evolution of the ordered compounds will require additional experimental techniques
like synchrotron XRD with \insitu{} heating to follow more precisely the evolution of the matrix and compounds lattice parameters,
and hence of their composition, during the different heat treatments. 
\Insitu{} heating experiments in a TEM are also appealing to image the evolution of these precipitates,
in particular of their size, shape, and density.
Once the thermal conditions have been found to get closer to equilibrium without dissolving the precipitates, 
it should be possible to use analytical TEM or atom probe tomography to check that equilibrium precipitates
have a composition close to Ti$_6$O stoichiometry, as expected from the equilibrium phase diagram \cite{Gunda2018}.

\section{Conclusions}

\Abinitio{} calculations correctly predict an increase of titanium lattice parameters
with the oxygen concentration for compositions between pure Ti and Ti$_6$O.
The increase is more important for $c$ than for $a$ parameter, in agreement with experiments.
The ordering state of oxygen atoms in the octahedral sites of the hcp lattice
has an influence on these lattice parameters, but the impact is less important 
than the one of oxygen concentration: for a same oxygen composition, 
an ordered compound has a slightly larger (respectively lower) $c$ parameter
(respectively $a$ parameter).
This is true both at 0\,K and at finite temperature, when lattice expansion 
is considered in the \abinitio{} approach through the quasi-harmonic approximation.

Knowing the impact of both oxygen concentration and oxygen ordering on lattice parameters, 
it is theoretically possible to relate the lattice mismatches of ordered compounds existing in Ti-O binary alloys
to their composition. 
Although the approach cannot be fully quantitative, because of the limited accuracy of \abinitio{} phonon calculations 
and their difficulty to predict small variations of lattice parameters, 
it nevertheless leads to new information on the composition of the different phases.
Applying the approach to an alloy containing 6000\,wppm O ($\xO=1.81$\%), 
one can conclude that the ordered compounds, which are present after a heat treatment at 1023\,K
followed by a water quench, do not have the expected Ti$_6$O stoichiometry
but a composition close to the alloy nominal concentration. 
This makes these precipitates undetectable by experimental techniques 
which are sensitive only to chemistry, like analytical TEM or atom probe tomography.
Oxygen ordering proceeds therefore at a faster pace than partitioning in this alloy: 
these precipitates have a long-range order corresponding to the Ti$_6$O structure,
as has been previously shown with TEM \cite{Poulain2022}, but their composition is lower 
than the equilibrium one corresponding to Ti$_6$O stoichiometry.
These precipitates, which are out of equilibrium and probably appear during the quench following the recrystallization heat treatment, 
do not evolve until reaching temperatures high enough to allow for significant diffusion of oxygen, 
\ie{} above 723\,K. 
It remains to be found in more details how the thermo-mechanical processing, in particular the temperature and duration of the annealing
or the cooling speed,
impact the composition of these ordered compounds.

\clearpage
\appendix
\section{Lattice expansion of pure Ti}
\label{sec:pureTi}

\begin{figure}[!b]
	\centering
	\includegraphics[width=0.7\linewidth]{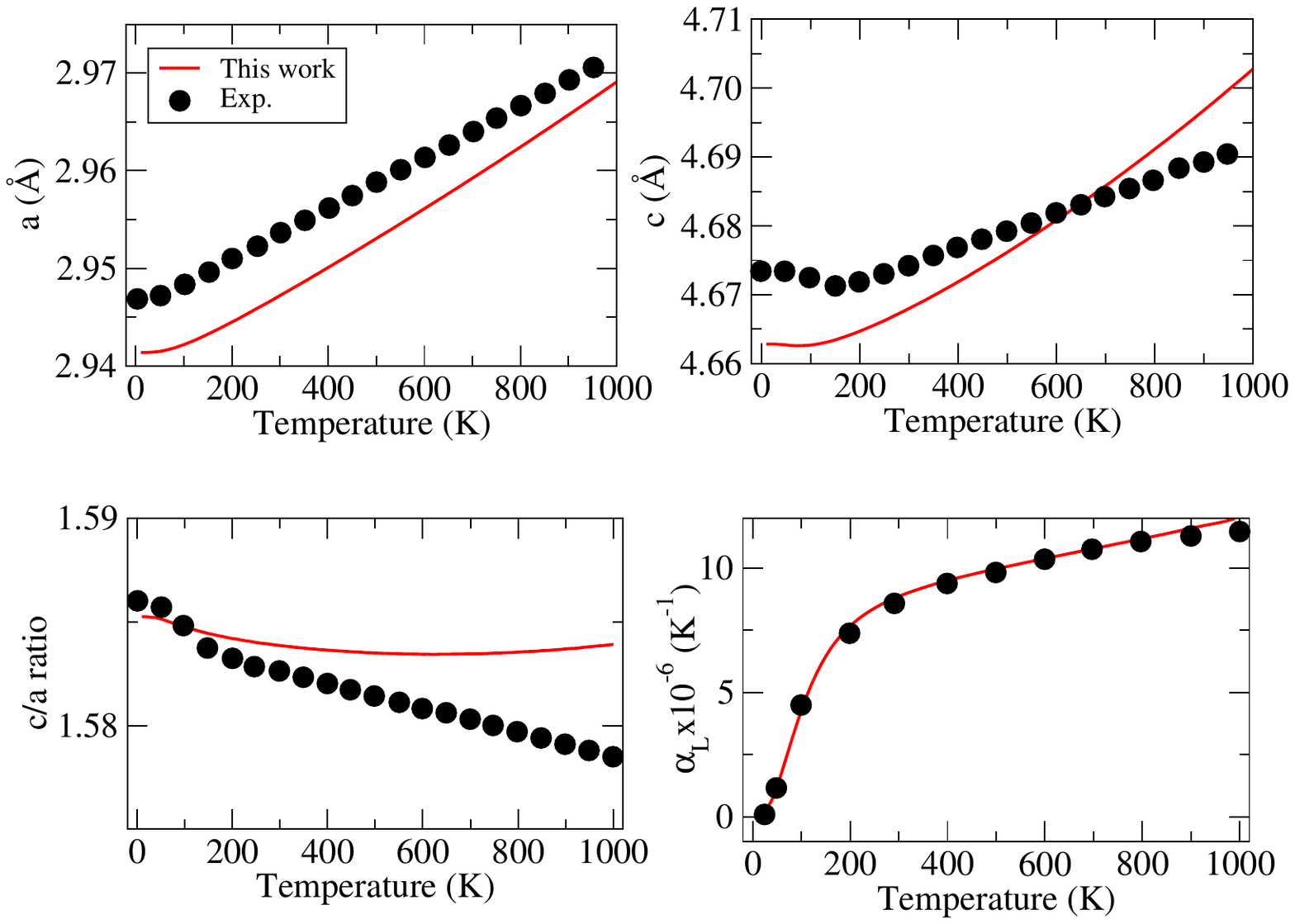}
	\caption{Lattice expansion of pure titanium as predicted by \abinitio{} calculations
	with the quasi-harmonic approximation (lines) and compared to the experimental data
	using recommended values of Touloukian \etal{} \cite{Touloukian1975}.
	}
	\label{fig:pureTi}
\end{figure}

We examine in this appendix how quantitative is our modeling approach relying on \abinitio{} calculations
and the quasi-harmonic approximation to predict the variations with temperature of the lattice parameters, 
considering pure Ti for which experimental data can be found \cite{Touloukian1975}.
Calculations have been performed in a supercell containing 96 Ti atoms 
which is a $4\times4\times3$ replication of the conventional hcp unit cell.
As shown in Fig. \ref{fig:pureTi}, the model reproduces reasonably well the experimental data,
in particular the anisotropy of the linear expansion characterized by a decreasing $c/a$ ratio with temperature
and a contraction of $c$ parameter at very low temperature,
in agreement with previous \abinitio{} calculations
\cite{Souvatzis2007,Mei2009,Argaman2016}.
The best agreement is found for the variations of the volume $V$ with the temperature, 
as characterized by the average linear expansion
\begin{equation*}
	\alpha_L = \frac{1}{3V} \left. \frac{\partial V}{\partial T} \right|_P .
\end{equation*}

\section{Impact of exchange-correlation functional}
\label{sec:LDA}

\begin{figure}[!bh]
	\centering
	\includegraphics[width=0.49\linewidth]{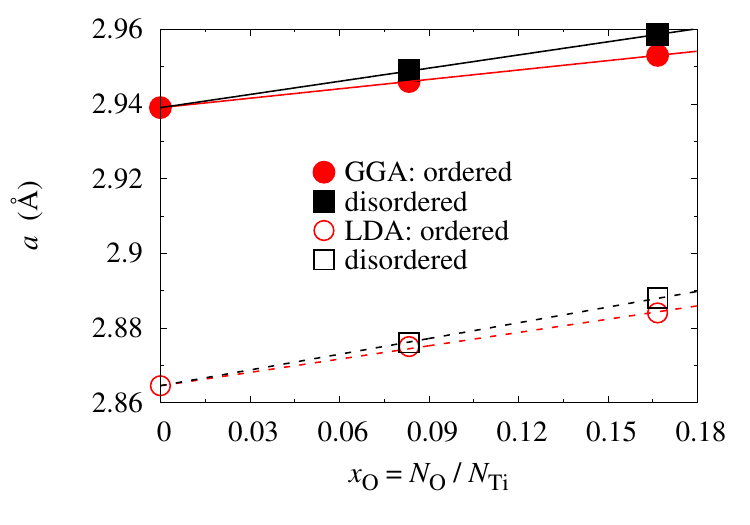}
	\hfill
	\includegraphics[width=0.49\linewidth]{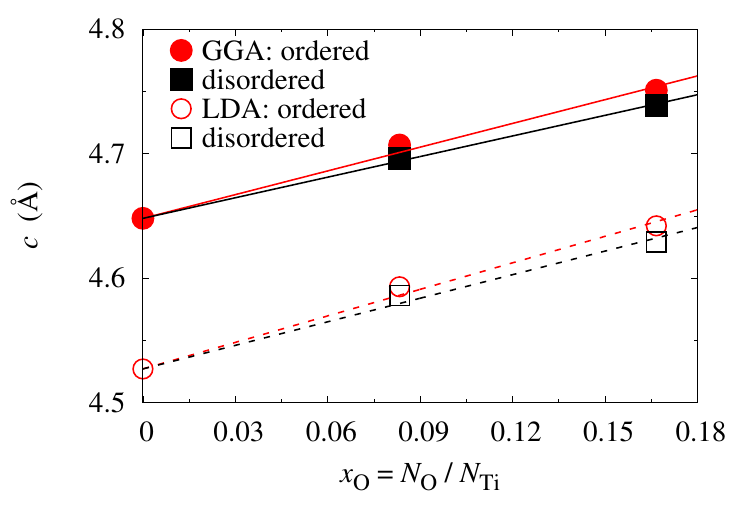}
	\caption{Variations of $a$ and $c$ lattice parameters with O concentration
	for ordered and disordered structures calculated at 0\,K 
	with two different exchange-correlation functionals, 
	GGA-PBE (full symbols continuous lines) and LDA (empty symbols, dashed lines).
	Lines are fit of the \abinitio{} calculations, imposing the lattice parameters of pure Ti
	for $x_{\rm O}=0$.}
	\label{fig:lattice_LDA}
\end{figure}

Equilibrium lattice parameters obtained from \abinitio{} calculations are known to depend
on the functional used to describe the electronic exchange and correlation.  
To test how the exchange-correlation functional impacts the obtained variations 
of the lattice parameters, we have performed additional calculations using the local density approximation (LDA)
with the parameterization of Perdew and Zunger \cite{Perdew1981}.
Except for the exchange-correlation functional, the \abinitio{} parameters are the same as for GGA calculations.
Calculations have been performed at 0\,K, neglecting atomic vibrations, for pure Ti, Ti$_{12}$O, and Ti$_6$O,
considering a disordered (SQS) and an ordered state. The structures, which rely on a 72 atoms supercell,
are the sames as the ones described in section \ref{sec:finiteT:methods}.
As expected, the LDA functional leads to slightly smaller lattice parameters than GGA (Fig. \ref{fig:lattice_LDA}).
But looking now to relative variations, one sees that LDA and GGA lead to the same linear variations
with the oxygen content. The lines describing these linear variations are parallel, with only an offset 
corresponding to the different lattice parameters predicted by LDA and GGA for pure Ti. 
This is true both for the ordered and disordered states, with the increase rate of $a$ (respectively $c$) 
lattice parameter being slightly higher (respectively lower) for the disordered than for the ordered state.
As long as one works with the lattice increases $\delta a$ and $\delta c$ induced by oxygen, 
there is thus no impact of the exchange-correlation functional used for \abinitio{} calculations,
at least when considering LDA and GGA-PBE functionals.

\clearpage

\section{XRD data}
\label{sec:XRD}

\begin{figure}[!h]
	\centering
	\includegraphics[width=0.7\linewidth]{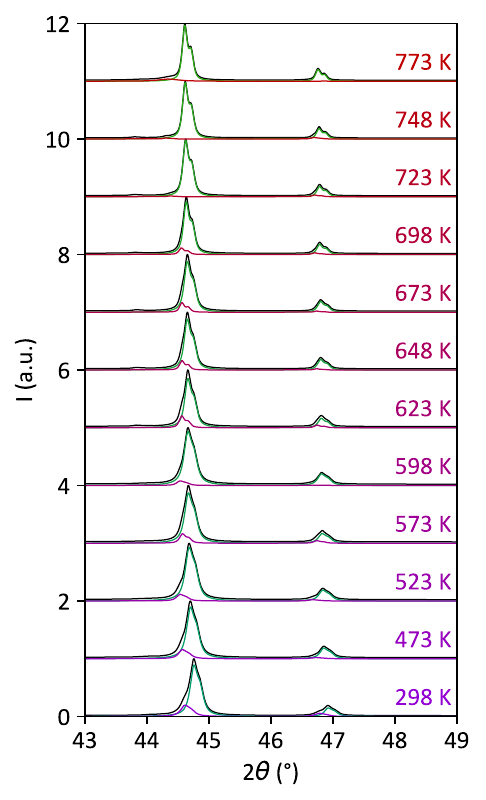}
	\caption{X-ray diffraction patterns obtained at different temperatures in the Ti-O alloy. 
	For each temperature, the experimental diffractogram (black) is shown 
	with the contributions of the $\alpha$-Ti matrix (green) and the ordered compounds (purple)
	obtained from Rietveld refinement.}
	\label{fig:XRD_all}
\end{figure}

\begin{acknowledgments}
	This work was performed using HPC resources from GENCI-IDRIS and -TGCC (Grants 2022-096847).
	The French National Research Agency (ANR) is acknowledged for the funding of TiTol Project [ANR-19-CE08-0032].
	Y. Millet from the Timet company is gratefully acknowledged for providing the alloys studied in this work.
\end{acknowledgments}

\bibliography{talla2023}

%apsrev4-2.bst 2019-01-14 (MD) hand-edited version of apsrev4-1.bst
%Control: key (0)
%Control: author (8) initials jnrlst
%Control: editor formatted (1) identically to author
%Control: production of article title (0) allowed
%Control: page (0) single
%Control: year (1) truncated
%Control: production of eprint (0) enabled
\begin{thebibliography}{39}%
\makeatletter
\providecommand \@ifxundefined [1]{%
 \@ifx{#1\undefined}
}%
\providecommand \@ifnum [1]{%
 \ifnum #1\expandafter \@firstoftwo
 \else \expandafter \@secondoftwo
 \fi
}%
\providecommand \@ifx [1]{%
 \ifx #1\expandafter \@firstoftwo
 \else \expandafter \@secondoftwo
 \fi
}%
\providecommand \natexlab [1]{#1}%
\providecommand \enquote  [1]{``#1''}%
\providecommand \bibnamefont  [1]{#1}%
\providecommand \bibfnamefont [1]{#1}%
\providecommand \citenamefont [1]{#1}%
\providecommand \href@noop [0]{\@secondoftwo}%
\providecommand \href [0]{\begingroup \@sanitize@url \@href}%
\providecommand \@href[1]{\@@startlink{#1}\@@href}%
\providecommand \@@href[1]{\endgroup#1\@@endlink}%
\providecommand \@sanitize@url [0]{\catcode `\\12\catcode `\$12\catcode
  `\&12\catcode `\#12\catcode `\^12\catcode `\_12\catcode `\%12\relax}%
\providecommand \@@startlink[1]{}%
\providecommand \@@endlink[0]{}%
\providecommand \url  [0]{\begingroup\@sanitize@url \@url }%
\providecommand \@url [1]{\endgroup\@href {#1}{\urlprefix }}%
\providecommand \urlprefix  [0]{URL }%
\providecommand \Eprint [0]{\href }%
\providecommand \doibase [0]{https://doi.org/}%
\providecommand \selectlanguage [0]{\@gobble}%
\providecommand \bibinfo  [0]{\@secondoftwo}%
\providecommand \bibfield  [0]{\@secondoftwo}%
\providecommand \translation [1]{[#1]}%
\providecommand \BibitemOpen [0]{}%
\providecommand \bibitemStop [0]{}%
\providecommand \bibitemNoStop [0]{.\EOS\space}%
\providecommand \EOS [0]{\spacefactor3000\relax}%
\providecommand \BibitemShut  [1]{\csname bibitem#1\endcsname}%
\let\auto@bib@innerbib\@empty
%</preamble>
\bibitem [{\citenamefont {Conrad}(1981)}]{Conrad1981}%
  \BibitemOpen
  \bibfield  {author} {\bibinfo {author} {\bibfnamefont {H.}~\bibnamefont
  {Conrad}},\ }\bibfield  {title} {\bibinfo {title} {Effect of interstitial
  solutes on the strength and ductility of titanium},\ }\href
  {https://doi.org/10.1016/0079-6425(81)90001-3} {\bibfield  {journal}
  {\bibinfo  {journal} {Prog. Mater. Sci.}\ }\textbf {\bibinfo {volume} {26}},\
  \bibinfo {pages} {123} (\bibinfo {year} {1981})}\BibitemShut {NoStop}%
\bibitem [{\citenamefont {Murray}\ and\ \citenamefont
  {Wriedt}(1987)}]{Murray1987}%
  \BibitemOpen
  \bibfield  {author} {\bibinfo {author} {\bibfnamefont {J.~L.}\ \bibnamefont
  {Murray}}\ and\ \bibinfo {author} {\bibfnamefont {H.~A.}\ \bibnamefont
  {Wriedt}},\ }\bibfield  {title} {\bibinfo {title} {The {O}-{T}i
  (oxygen-titanium) system},\ }\href {https://doi.org/10.1007/bf02873201}
  {\bibfield  {journal} {\bibinfo  {journal} {Bull. Alloy Phase Diagr.}\
  }\textbf {\bibinfo {volume} {8}},\ \bibinfo {pages} {148} (\bibinfo {year}
  {1987})}\BibitemShut {NoStop}%
\bibitem [{\citenamefont {Kornilov}\ and\ \citenamefont
  {Glazova}(1965)}]{Kornilov1965}%
  \BibitemOpen
  \bibfield  {author} {\bibinfo {author} {\bibfnamefont {I.~I.}\ \bibnamefont
  {Kornilov}}\ and\ \bibinfo {author} {\bibfnamefont {V.~V.}\ \bibnamefont
  {Glazova}},\ }\bibfield  {title} {\bibinfo {title} {Phase diagram of the
  system {T}i-{O}$_2$ and some properties of the alloys of this system},\ }in\
  \href@noop {} {\emph {\bibinfo {booktitle} {Physical metallurgy of
  titanium}}},\ \bibinfo {editor} {edited by\ \bibinfo {editor} {\bibfnamefont
  {I.~I.}\ \bibnamefont {Kornilov}}}\ (\bibinfo {year} {1965})\ pp.\ \bibinfo
  {pages} {12--28}\BibitemShut {NoStop}%
\bibitem [{\citenamefont {Yamaguchi}\ \emph {et~al.}(1966)\citenamefont
  {Yamaguchi}, \citenamefont {Koiwa},\ and\ \citenamefont
  {Hirabayashi}}]{Yamaguchi1966}%
  \BibitemOpen
  \bibfield  {author} {\bibinfo {author} {\bibfnamefont {S.}~\bibnamefont
  {Yamaguchi}}, \bibinfo {author} {\bibfnamefont {M.}~\bibnamefont {Koiwa}},\
  and\ \bibinfo {author} {\bibfnamefont {M.}~\bibnamefont {Hirabayashi}},\
  }\bibfield  {title} {\bibinfo {title} {Interstitial superlattice of
  {T}i$_6${O} and its transformation},\ }\href
  {https://doi.org/10.1143/jpsj.21.2096} {\bibfield  {journal} {\bibinfo
  {journal} {J. Phys. Soc. Jpn.}\ }\textbf {\bibinfo {volume} {21}},\ \bibinfo
  {pages} {2096} (\bibinfo {year} {1966})}\BibitemShut {NoStop}%
\bibitem [{\citenamefont {Yamaguchi}(1969)}]{Yamaguchi1969}%
  \BibitemOpen
  \bibfield  {author} {\bibinfo {author} {\bibfnamefont {S.}~\bibnamefont
  {Yamaguchi}},\ }\bibfield  {title} {\bibinfo {title} {Interstitial
  order-disorder transformation in the {T}i-{O} solid solution. {I}. {O}rdered
  arrangement of oxygen},\ }\href {https://doi.org/10.1143/jpsj.27.155}
  {\bibfield  {journal} {\bibinfo  {journal} {J. Phys. Soc. Jpn.}\ }\textbf
  {\bibinfo {volume} {27}},\ \bibinfo {pages} {155} (\bibinfo {year}
  {1969})}\BibitemShut {NoStop}%
\bibitem [{\citenamefont {Hirabayashi}\ \emph {et~al.}(1974)\citenamefont
  {Hirabayashi}, \citenamefont {Yamaguchi}, \citenamefont {Asano},\ and\
  \citenamefont {Hiraga}}]{Hirabayashi1974}%
  \BibitemOpen
  \bibfield  {author} {\bibinfo {author} {\bibfnamefont {M.}~\bibnamefont
  {Hirabayashi}}, \bibinfo {author} {\bibfnamefont {S.}~\bibnamefont
  {Yamaguchi}}, \bibinfo {author} {\bibfnamefont {H.}~\bibnamefont {Asano}},\
  and\ \bibinfo {author} {\bibfnamefont {K.}~\bibnamefont {Hiraga}},\
  }\bibfield  {title} {\bibinfo {title} {Order-disorder transformations of
  interstitial solutes in transition metals of {IV} and {V} groups},\ }in\
  \href {https://doi.org/10.1007/978-3-642-80840-1_10} {\emph {\bibinfo
  {booktitle} {Order-Disorder Transformations in Alloys}}}\ (\bibinfo
  {publisher} {Springer},\ \bibinfo {address} {Berlin Heidelberg},\ \bibinfo
  {year} {1974})\ pp.\ \bibinfo {pages} {266--302}\BibitemShut {NoStop}%
\bibitem [{\citenamefont {Banerjee}\ and\ \citenamefont
  {Mukhopadhyay}(2007)}]{Banerjee2007}%
  \BibitemOpen
  \bibfield  {author} {\bibinfo {author} {\bibfnamefont {S.}~\bibnamefont
  {Banerjee}}\ and\ \bibinfo {author} {\bibfnamefont {P.}~\bibnamefont
  {Mukhopadhyay}},\ }\href
  {https://www.sciencedirect.com/science/bookseries/14701804/12} {\emph
  {\bibinfo {title} {Phase transformations: examples from titanium and
  zirconium alloys}}},\ edited by\ \bibinfo {editor} {\bibfnamefont {R.~W.}\
  \bibnamefont {Cahn}},\ \bibinfo {series} {Pergamon Materials Series},
  Vol.~\bibinfo {volume} {12}\ (\bibinfo  {publisher} {Elsevier Science},\
  \bibinfo {year} {2007})\BibitemShut {NoStop}%
\bibitem [{\citenamefont {Ruban}\ \emph {et~al.}(2010)\citenamefont {Ruban},
  \citenamefont {Baykov}, \citenamefont {Johansson}, \citenamefont {Dmitriev},\
  and\ \citenamefont {Blanter}}]{Ruban2010}%
  \BibitemOpen
  \bibfield  {author} {\bibinfo {author} {\bibfnamefont {A.~V.}\ \bibnamefont
  {Ruban}}, \bibinfo {author} {\bibfnamefont {V.~I.}\ \bibnamefont {Baykov}},
  \bibinfo {author} {\bibfnamefont {B.}~\bibnamefont {Johansson}}, \bibinfo
  {author} {\bibfnamefont {V.~V.}\ \bibnamefont {Dmitriev}},\ and\ \bibinfo
  {author} {\bibfnamefont {M.~S.}\ \bibnamefont {Blanter}},\ }\bibfield
  {title} {\bibinfo {title} {Oxygen and nitrogen interstitial ordering in hcp
  {T}i, {Z}r, and {H}f: An ab initio study},\ }\href
  {https://doi.org/10.1103/physrevb.82.134110} {\bibfield  {journal} {\bibinfo
  {journal} {Phys. Rev. B}\ }\textbf {\bibinfo {volume} {82}},\ \bibinfo
  {pages} {134110} (\bibinfo {year} {2010})}\BibitemShut {NoStop}%
\bibitem [{\citenamefont {Burton}\ and\ \citenamefont {van~de
  Walle}(2012)}]{Burton2012c}%
  \BibitemOpen
  \bibfield  {author} {\bibinfo {author} {\bibfnamefont {B.~P.}\ \bibnamefont
  {Burton}}\ and\ \bibinfo {author} {\bibfnamefont {A.}~\bibnamefont {van~de
  Walle}},\ }\bibfield  {title} {\bibinfo {title} {First principles phase
  diagram calculations for the octahedral-interstitial system
  $\alpha${TiO}$_x$, 0 $\leq x \leq1/2$},\ }\href
  {https://doi.org/10.1016/j.calphad.2012.09.004} {\bibfield  {journal}
  {\bibinfo  {journal} {Calphad}\ }\textbf {\bibinfo {volume} {39}},\ \bibinfo
  {pages} {97} (\bibinfo {year} {2012})}\BibitemShut {NoStop}%
\bibitem [{\citenamefont {Gunda}\ \emph {et~al.}(2018)\citenamefont {Gunda},
  \citenamefont {Puchala},\ and\ \citenamefont {der Ven}}]{Gunda2018}%
  \BibitemOpen
  \bibfield  {author} {\bibinfo {author} {\bibfnamefont {N.~S.~H.}\
  \bibnamefont {Gunda}}, \bibinfo {author} {\bibfnamefont {B.}~\bibnamefont
  {Puchala}},\ and\ \bibinfo {author} {\bibfnamefont {A.~V.}\ \bibnamefont {der
  Ven}},\ }\bibfield  {title} {\bibinfo {title} {Resolving phase stability in
  the {T}i-{O} binary with first-principles statistical mechanics methods},\
  }\href {https://doi.org/10.1103/physrevmaterials.2.033604} {\bibfield
  {journal} {\bibinfo  {journal} {Phys. Rev. Mater.}\ }\textbf {\bibinfo
  {volume} {2}},\ \bibinfo {pages} {033604} (\bibinfo {year}
  {2018})}\BibitemShut {NoStop}%
\bibitem [{\citenamefont {Yu}\ \emph {et~al.}(2015)\citenamefont {Yu},
  \citenamefont {Qi}, \citenamefont {Tsuru}, \citenamefont {Traylor},
  \citenamefont {Rugg}, \citenamefont {Morris}, \citenamefont {Asta},
  \citenamefont {Chrzan},\ and\ \citenamefont {Minor}}]{Yu2015}%
  \BibitemOpen
  \bibfield  {author} {\bibinfo {author} {\bibfnamefont {Q.}~\bibnamefont
  {Yu}}, \bibinfo {author} {\bibfnamefont {L.}~\bibnamefont {Qi}}, \bibinfo
  {author} {\bibfnamefont {T.}~\bibnamefont {Tsuru}}, \bibinfo {author}
  {\bibfnamefont {R.}~\bibnamefont {Traylor}}, \bibinfo {author} {\bibfnamefont
  {D.}~\bibnamefont {Rugg}}, \bibinfo {author} {\bibfnamefont {J.~W.}\
  \bibnamefont {Morris}}, \bibinfo {author} {\bibfnamefont {M.}~\bibnamefont
  {Asta}}, \bibinfo {author} {\bibfnamefont {D.~C.}\ \bibnamefont {Chrzan}},\
  and\ \bibinfo {author} {\bibfnamefont {A.~M.}\ \bibnamefont {Minor}},\
  }\bibfield  {title} {\bibinfo {title} {Origin of dramatic oxygen solute
  strengthening effect in titanium},\ }\href
  {https://doi.org/10.1126/science.1260485} {\bibfield  {journal} {\bibinfo
  {journal} {Science}\ }\textbf {\bibinfo {volume} {347}},\ \bibinfo {pages}
  {635} (\bibinfo {year} {2015})}\BibitemShut {NoStop}%
\bibitem [{\citenamefont {Barkia}\ \emph {et~al.}(2015)\citenamefont {Barkia},
  \citenamefont {Doquet}, \citenamefont {Couzinié}, \citenamefont {Guillot},\
  and\ \citenamefont {Héripré}}]{Barkia2015}%
  \BibitemOpen
  \bibfield  {author} {\bibinfo {author} {\bibfnamefont {B.}~\bibnamefont
  {Barkia}}, \bibinfo {author} {\bibfnamefont {V.}~\bibnamefont {Doquet}},
  \bibinfo {author} {\bibfnamefont {J.-P.}\ \bibnamefont {Couzinié}}, \bibinfo
  {author} {\bibfnamefont {I.}~\bibnamefont {Guillot}},\ and\ \bibinfo {author}
  {\bibfnamefont {E.}~\bibnamefont {Héripré}},\ }\bibfield  {title} {\bibinfo
  {title} {In situ monitoring of the deformation mechanisms in titanium with
  different oxygen contents},\ }\href
  {https://doi.org/10.1016/j.msea.2015.03.044} {\bibfield  {journal} {\bibinfo
  {journal} {Mater. Sci. Eng., A}\ }\textbf {\bibinfo {volume} {636}},\
  \bibinfo {pages} {91} (\bibinfo {year} {2015})}\BibitemShut {NoStop}%
\bibitem [{\citenamefont {Barkia}\ \emph {et~al.}(2017)\citenamefont {Barkia},
  \citenamefont {Couzinié}, \citenamefont {Lartigue-Korinek}, \citenamefont
  {Guillot},\ and\ \citenamefont {Doquet}}]{Barkia2017}%
  \BibitemOpen
  \bibfield  {author} {\bibinfo {author} {\bibfnamefont {B.}~\bibnamefont
  {Barkia}}, \bibinfo {author} {\bibfnamefont {J.-P.}\ \bibnamefont
  {Couzinié}}, \bibinfo {author} {\bibfnamefont {S.}~\bibnamefont
  {Lartigue-Korinek}}, \bibinfo {author} {\bibfnamefont {I.}~\bibnamefont
  {Guillot}},\ and\ \bibinfo {author} {\bibfnamefont {V.}~\bibnamefont
  {Doquet}},\ }\bibfield  {title} {\bibinfo {title} {In situ {TEM} observations
  of dislocation dynamics in $\alpha$ titanium: Effect of the oxygen content},\
  }\href {https://doi.org/10.1016/j.msea.2017.07.040} {\bibfield  {journal}
  {\bibinfo  {journal} {Mater. Sci. Eng., A}\ }\textbf {\bibinfo {volume}
  {703}},\ \bibinfo {pages} {331} (\bibinfo {year} {2017})}\BibitemShut
  {NoStop}%
\bibitem [{\citenamefont {Chaari}\ \emph {et~al.}(2019)\citenamefont {Chaari},
  \citenamefont {Rodney},\ and\ \citenamefont {Clouet}}]{Chaari2019}%
  \BibitemOpen
  \bibfield  {author} {\bibinfo {author} {\bibfnamefont {N.}~\bibnamefont
  {Chaari}}, \bibinfo {author} {\bibfnamefont {D.}~\bibnamefont {Rodney}},\
  and\ \bibinfo {author} {\bibfnamefont {E.}~\bibnamefont {Clouet}},\
  }\bibfield  {title} {\bibinfo {title} {Oxygen-dislocation interaction in
  titanium from first principles},\ }\href
  {https://doi.org/10.1016/j.scriptamat.2018.11.025} {\bibfield  {journal}
  {\bibinfo  {journal} {Scr. Mater.}\ }\textbf {\bibinfo {volume} {162}},\
  \bibinfo {pages} {200} (\bibinfo {year} {2019})}\BibitemShut {NoStop}%
\bibitem [{\citenamefont {Chong}\ \emph {et~al.}(2020)\citenamefont {Chong},
  \citenamefont {Poschmann}, \citenamefont {Zhang}, \citenamefont {Zhao},
  \citenamefont {Hooshmand}, \citenamefont {Rothchild}, \citenamefont
  {Olmsted}, \citenamefont {Morris}, \citenamefont {Chrzan}, \citenamefont
  {Asta},\ and\ \citenamefont {Minor}}]{Chong2020}%
  \BibitemOpen
  \bibfield  {author} {\bibinfo {author} {\bibfnamefont {Y.}~\bibnamefont
  {Chong}}, \bibinfo {author} {\bibfnamefont {M.}~\bibnamefont {Poschmann}},
  \bibinfo {author} {\bibfnamefont {R.}~\bibnamefont {Zhang}}, \bibinfo
  {author} {\bibfnamefont {S.}~\bibnamefont {Zhao}}, \bibinfo {author}
  {\bibfnamefont {M.~S.}\ \bibnamefont {Hooshmand}}, \bibinfo {author}
  {\bibfnamefont {E.}~\bibnamefont {Rothchild}}, \bibinfo {author}
  {\bibfnamefont {D.~L.}\ \bibnamefont {Olmsted}}, \bibinfo {author}
  {\bibfnamefont {J.~W.}\ \bibnamefont {Morris}}, \bibinfo {author}
  {\bibfnamefont {D.~C.}\ \bibnamefont {Chrzan}}, \bibinfo {author}
  {\bibfnamefont {M.}~\bibnamefont {Asta}},\ and\ \bibinfo {author}
  {\bibfnamefont {A.~M.}\ \bibnamefont {Minor}},\ }\bibfield  {title} {\bibinfo
  {title} {Mechanistic basis of oxygen sensitivity in titanium},\ }\href
  {https://doi.org/10.1126/sciadv.abc4060} {\bibfield  {journal} {\bibinfo
  {journal} {Sci. Adv.}\ }\textbf {\bibinfo {volume} {6}},\ \bibinfo {pages}
  {eabc4060} (\bibinfo {year} {2020})}\BibitemShut {NoStop}%
\bibitem [{\citenamefont {Chong}\ \emph {et~al.}(2023)\citenamefont {Chong},
  \citenamefont {Gholizadeh}, \citenamefont {Tsuru}, \citenamefont {Zhang},
  \citenamefont {Inoue}, \citenamefont {Gao}, \citenamefont {Godfrey},
  \citenamefont {Mitsuhara}, \citenamefont {Morris}, \citenamefont {Minor},\
  and\ \citenamefont {Tsuji}}]{Chong2023}%
  \BibitemOpen
  \bibfield  {author} {\bibinfo {author} {\bibfnamefont {Y.}~\bibnamefont
  {Chong}}, \bibinfo {author} {\bibfnamefont {R.}~\bibnamefont {Gholizadeh}},
  \bibinfo {author} {\bibfnamefont {T.}~\bibnamefont {Tsuru}}, \bibinfo
  {author} {\bibfnamefont {R.}~\bibnamefont {Zhang}}, \bibinfo {author}
  {\bibfnamefont {K.}~\bibnamefont {Inoue}}, \bibinfo {author} {\bibfnamefont
  {W.}~\bibnamefont {Gao}}, \bibinfo {author} {\bibfnamefont {A.}~\bibnamefont
  {Godfrey}}, \bibinfo {author} {\bibfnamefont {M.}~\bibnamefont {Mitsuhara}},
  \bibinfo {author} {\bibfnamefont {J.~W.}\ \bibnamefont {Morris}}, \bibinfo
  {author} {\bibfnamefont {A.~M.}\ \bibnamefont {Minor}},\ and\ \bibinfo
  {author} {\bibfnamefont {N.}~\bibnamefont {Tsuji}},\ }\bibfield  {title}
  {\bibinfo {title} {Grain refinement in titanium prevents low temperature
  oxygen embrittlement},\ }\href {https://doi.org/10.1038/s41467-023-36030-0}
  {\bibfield  {journal} {\bibinfo  {journal} {Nat. Commun.}\ }\textbf {\bibinfo
  {volume} {14}},\ \bibinfo {pages} {404} (\bibinfo {year} {2023})}\BibitemShut
  {NoStop}%
\bibitem [{\citenamefont {Poulain}\ \emph {et~al.}(2022)\citenamefont
  {Poulain}, \citenamefont {Delannoy}, \citenamefont {Guillot}, \citenamefont
  {Amann}, \citenamefont {Guillou}, \citenamefont {Lartigue-Korinek},
  \citenamefont {Thiaudière}, \citenamefont {Béchade}, \citenamefont
  {Clouet},\ and\ \citenamefont {Prima}}]{Poulain2022}%
  \BibitemOpen
  \bibfield  {author} {\bibinfo {author} {\bibfnamefont {R.}~\bibnamefont
  {Poulain}}, \bibinfo {author} {\bibfnamefont {S.}~\bibnamefont {Delannoy}},
  \bibinfo {author} {\bibfnamefont {I.}~\bibnamefont {Guillot}}, \bibinfo
  {author} {\bibfnamefont {F.}~\bibnamefont {Amann}}, \bibinfo {author}
  {\bibfnamefont {R.}~\bibnamefont {Guillou}}, \bibinfo {author} {\bibfnamefont
  {S.}~\bibnamefont {Lartigue-Korinek}}, \bibinfo {author} {\bibfnamefont
  {D.}~\bibnamefont {Thiaudière}}, \bibinfo {author} {\bibfnamefont {J.-L.}\
  \bibnamefont {Béchade}}, \bibinfo {author} {\bibfnamefont {E.}~\bibnamefont
  {Clouet}},\ and\ \bibinfo {author} {\bibfnamefont {F.}~\bibnamefont
  {Prima}},\ }\bibfield  {title} {\bibinfo {title} {First experimental evidence
  of oxygen ordering in dilute titanium{\textendash}oxygen alloys},\ }\href
  {https://doi.org/10.1080/21663831.2022.2057202} {\bibfield  {journal}
  {\bibinfo  {journal} {Mater. Res. Lett.}\ }\textbf {\bibinfo {volume} {10}},\
  \bibinfo {pages} {481} (\bibinfo {year} {2022})}\BibitemShut {NoStop}%
\bibitem [{\citenamefont {Kornilov}(1973)}]{Kornilov1973}%
  \BibitemOpen
  \bibfield  {author} {\bibinfo {author} {\bibfnamefont {I.~I.}\ \bibnamefont
  {Kornilov}},\ }\bibfield  {title} {\bibinfo {title} {Effect of oxygen on
  titanium and its alloys},\ }\href {https://doi.org/10.1007/bf00656056}
  {\bibfield  {journal} {\bibinfo  {journal} {Met. Sci. Heat Treat.}\ }\textbf
  {\bibinfo {volume} {15}},\ \bibinfo {pages} {826} (\bibinfo {year}
  {1973})}\BibitemShut {NoStop}%
\bibitem [{\citenamefont {Amann}\ \emph
  {et~al.}(2023{\natexlab{a}})\citenamefont {Amann}, \citenamefont {Poulain},
  \citenamefont {Delannoy}, \citenamefont {Couzinié}, \citenamefont {Clouet},
  \citenamefont {Guillot},\ and\ \citenamefont {Prima}}]{Amann2023}%
  \BibitemOpen
  \bibfield  {author} {\bibinfo {author} {\bibfnamefont {F.}~\bibnamefont
  {Amann}}, \bibinfo {author} {\bibfnamefont {R.}~\bibnamefont {Poulain}},
  \bibinfo {author} {\bibfnamefont {S.}~\bibnamefont {Delannoy}}, \bibinfo
  {author} {\bibfnamefont {J.-P.}\ \bibnamefont {Couzinié}}, \bibinfo {author}
  {\bibfnamefont {E.}~\bibnamefont {Clouet}}, \bibinfo {author} {\bibfnamefont
  {I.}~\bibnamefont {Guillot}},\ and\ \bibinfo {author} {\bibfnamefont
  {F.}~\bibnamefont {Prima}},\ }\bibfield  {title} {\bibinfo {title} {An
  improved combination of tensile strength and ductility in titanium alloys via
  oxygen ordering},\ }\href {https://doi.org/10.1016/j.msea.2023.144720}
  {\bibfield  {journal} {\bibinfo  {journal} {Mater. Sci. Eng. A}\ }\textbf
  {\bibinfo {volume} {867}},\ \bibinfo {pages} {144720} (\bibinfo {year}
  {2023}{\natexlab{a}})}\BibitemShut {NoStop}%
\bibitem [{\citenamefont {Kresse}\ and\ \citenamefont
  {Furthmüller}(1996)}]{Kresse1996}%
  \BibitemOpen
  \bibfield  {author} {\bibinfo {author} {\bibfnamefont {G.}~\bibnamefont
  {Kresse}}\ and\ \bibinfo {author} {\bibfnamefont {J.}~\bibnamefont
  {Furthmüller}},\ }\bibfield  {title} {\bibinfo {title} {Efficiency of
  ab-initio total energy calculations for metals and semiconductors using a
  plane-wave basis set},\ }\href {https://doi.org/10.1016/0927-0256(96)00008-0}
  {\bibfield  {journal} {\bibinfo  {journal} {Comput. Mater. Sci.}\ }\textbf
  {\bibinfo {volume} {6}},\ \bibinfo {pages} {15} (\bibinfo {year}
  {1996})}\BibitemShut {NoStop}%
\bibitem [{\citenamefont {Perdew}\ \emph {et~al.}(1996)\citenamefont {Perdew},
  \citenamefont {Burke},\ and\ \citenamefont {Ernzerhof}}]{Perdew1996}%
  \BibitemOpen
  \bibfield  {author} {\bibinfo {author} {\bibfnamefont {J.~P.}\ \bibnamefont
  {Perdew}}, \bibinfo {author} {\bibfnamefont {K.}~\bibnamefont {Burke}},\ and\
  \bibinfo {author} {\bibfnamefont {M.}~\bibnamefont {Ernzerhof}},\ }\bibfield
  {title} {\bibinfo {title} {Generalized gradient approximation made simple},\
  }\href {https://doi.org/10.1103/PhysRevLett.77.3865} {\bibfield  {journal}
  {\bibinfo  {journal} {Phys. Rev. Lett.}\ }\textbf {\bibinfo {volume} {77}},\
  \bibinfo {pages} {3865} (\bibinfo {year} {1996})}\BibitemShut {NoStop}%
\bibitem [{\citenamefont {Blöchl}(1994)}]{Bloechl1994}%
  \BibitemOpen
  \bibfield  {author} {\bibinfo {author} {\bibfnamefont {P.~E.}\ \bibnamefont
  {Blöchl}},\ }\bibfield  {title} {\bibinfo {title} {Projector augmented-wave
  method},\ }\href {https://doi.org/10.1103/physrevb.50.17953} {\bibfield
  {journal} {\bibinfo  {journal} {Phys. Rev. B}\ }\textbf {\bibinfo {volume}
  {50}},\ \bibinfo {pages} {17953} (\bibinfo {year} {1994})}\BibitemShut
  {NoStop}%
\bibitem [{\citenamefont {Monkhorst}\ and\ \citenamefont
  {Pack}(1976)}]{Monkhorst1976}%
  \BibitemOpen
  \bibfield  {author} {\bibinfo {author} {\bibfnamefont {H.~J.}\ \bibnamefont
  {Monkhorst}}\ and\ \bibinfo {author} {\bibfnamefont {J.~D.}\ \bibnamefont
  {Pack}},\ }\bibfield  {title} {\bibinfo {title} {Special points for
  {B}rillouin-zone integrations},\ }\href
  {https://doi.org/10.1103/physrevb.13.5188} {\bibfield  {journal} {\bibinfo
  {journal} {Phys. Rev. B}\ }\textbf {\bibinfo {volume} {13}},\ \bibinfo
  {pages} {5188} (\bibinfo {year} {1976})}\BibitemShut {NoStop}%
\bibitem [{\citenamefont {van~de Walle}\ \emph {et~al.}(2013)\citenamefont
  {van~de Walle}, \citenamefont {Tiwary}, \citenamefont {de~Jong},
  \citenamefont {Olmsted}, \citenamefont {Asta}, \citenamefont {Dick},
  \citenamefont {Shin}, \citenamefont {Wang}, \citenamefont {Chen},\ and\
  \citenamefont {Liu}}]{vandeWalle2013}%
  \BibitemOpen
  \bibfield  {author} {\bibinfo {author} {\bibfnamefont {A.}~\bibnamefont
  {van~de Walle}}, \bibinfo {author} {\bibfnamefont {P.}~\bibnamefont
  {Tiwary}}, \bibinfo {author} {\bibfnamefont {M.}~\bibnamefont {de~Jong}},
  \bibinfo {author} {\bibfnamefont {D.~L.}\ \bibnamefont {Olmsted}}, \bibinfo
  {author} {\bibfnamefont {M.}~\bibnamefont {Asta}}, \bibinfo {author}
  {\bibfnamefont {A.}~\bibnamefont {Dick}}, \bibinfo {author} {\bibfnamefont
  {D.}~\bibnamefont {Shin}}, \bibinfo {author} {\bibfnamefont {Y.}~\bibnamefont
  {Wang}}, \bibinfo {author} {\bibfnamefont {L.-Q.}\ \bibnamefont {Chen}},\
  and\ \bibinfo {author} {\bibfnamefont {Z.-K.}\ \bibnamefont {Liu}},\
  }\bibfield  {title} {\bibinfo {title} {Efficient stochastic generation of
  special quasirandom structures},\ }\href
  {https://doi.org/10.1016/j.calphad.2013.06.006} {\bibfield  {journal}
  {\bibinfo  {journal} {Calphad}\ }\textbf {\bibinfo {volume} {42}},\ \bibinfo
  {pages} {13} (\bibinfo {year} {2013})}\BibitemShut {NoStop}%
\bibitem [{\citenamefont {Yamaguchi}\ \emph {et~al.}(1970)\citenamefont
  {Yamaguchi}, \citenamefont {Hiraga},\ and\ \citenamefont
  {Hirabayashi}}]{Yamaguchi1970}%
  \BibitemOpen
  \bibfield  {author} {\bibinfo {author} {\bibfnamefont {S.}~\bibnamefont
  {Yamaguchi}}, \bibinfo {author} {\bibfnamefont {K.}~\bibnamefont {Hiraga}},\
  and\ \bibinfo {author} {\bibfnamefont {M.}~\bibnamefont {Hirabayashi}},\
  }\bibfield  {title} {\bibinfo {title} {Interstitial order-disorder
  transformation in the {T}i-{O} solid solution. {IV}. {A} neutron diffraction
  study},\ }\href {https://doi.org/10.1143/jpsj.28.1014} {\bibfield  {journal}
  {\bibinfo  {journal} {J. Phys. Soc. Jpn.}\ }\textbf {\bibinfo {volume}
  {28}},\ \bibinfo {pages} {1014} (\bibinfo {year} {1970})}\BibitemShut
  {NoStop}%
\bibitem [{\citenamefont {Wiedemann}\ \emph {et~al.}(1987)\citenamefont
  {Wiedemann}, \citenamefont {Shenoy},\ and\ \citenamefont
  {Unnam}}]{Wiedemann1987}%
  \BibitemOpen
  \bibfield  {author} {\bibinfo {author} {\bibfnamefont {K.~E.}\ \bibnamefont
  {Wiedemann}}, \bibinfo {author} {\bibfnamefont {R.~N.}\ \bibnamefont
  {Shenoy}},\ and\ \bibinfo {author} {\bibfnamefont {J.}~\bibnamefont
  {Unnam}},\ }\bibfield  {title} {\bibinfo {title} {Microhardness and lattice
  parameter calibrations of the oxygen solid solutions of unalloyed
  $\alpha$-titanium and {T}i-6{A}l-2{S}n-4{Z}r-2{M}o},\ }\href
  {https://doi.org/10.1007/bf02646662} {\bibfield  {journal} {\bibinfo
  {journal} {Metall. Trans. A}\ }\textbf {\bibinfo {volume} {18}},\ \bibinfo
  {pages} {1503} (\bibinfo {year} {1987})}\BibitemShut {NoStop}%
\bibitem [{\citenamefont {Togo}\ and\ \citenamefont {Tanaka}(2015)}]{Togo2015}%
  \BibitemOpen
  \bibfield  {author} {\bibinfo {author} {\bibfnamefont {A.}~\bibnamefont
  {Togo}}\ and\ \bibinfo {author} {\bibfnamefont {I.}~\bibnamefont {Tanaka}},\
  }\bibfield  {title} {\bibinfo {title} {First principles phonon calculations
  in materials science},\ }\href
  {https://doi.org/10.1016/j.scriptamat.2015.07.021} {\bibfield  {journal}
  {\bibinfo  {journal} {Scripta Mater.}\ }\textbf {\bibinfo {volume} {108}},\
  \bibinfo {pages} {1} (\bibinfo {year} {2015})}\BibitemShut {NoStop}%
\bibitem [{\citenamefont {Wolverton}\ and\ \citenamefont
  {Zunger}(1995)}]{Wolverton1995}%
  \BibitemOpen
  \bibfield  {author} {\bibinfo {author} {\bibfnamefont {C.}~\bibnamefont
  {Wolverton}}\ and\ \bibinfo {author} {\bibfnamefont {A.}~\bibnamefont
  {Zunger}},\ }\bibfield  {title} {\bibinfo {title} {First-principles theory of
  short-range order, electronic excitations, and spin polarization in {N}i-{V}
  and {P}d-{V} alloys},\ }\href {https://doi.org/10.1103/PhysRevB.52.8813}
  {\bibfield  {journal} {\bibinfo  {journal} {Phys. Rev. B}\ }\textbf {\bibinfo
  {volume} {52}},\ \bibinfo {pages} {8813} (\bibinfo {year}
  {1995})}\BibitemShut {NoStop}%
\bibitem [{\citenamefont {Zhang}\ \emph {et~al.}(2017)\citenamefont {Zhang},
  \citenamefont {Grabowski}, \citenamefont {K\"ormann}, \citenamefont
  {Freysoldt},\ and\ \citenamefont {Neugebauer}}]{Zhang2017}%
  \BibitemOpen
  \bibfield  {author} {\bibinfo {author} {\bibfnamefont {X.}~\bibnamefont
  {Zhang}}, \bibinfo {author} {\bibfnamefont {B.}~\bibnamefont {Grabowski}},
  \bibinfo {author} {\bibfnamefont {F.}~\bibnamefont {K\"ormann}}, \bibinfo
  {author} {\bibfnamefont {C.}~\bibnamefont {Freysoldt}},\ and\ \bibinfo
  {author} {\bibfnamefont {J.}~\bibnamefont {Neugebauer}},\ }\bibfield  {title}
  {\bibinfo {title} {Accurate electronic free energies of the $3d$, $4d$, and
  $5d$ transition metals at high temperatures},\ }\href
  {https://doi.org/10.1103/PhysRevB.95.165126} {\bibfield  {journal} {\bibinfo
  {journal} {Phys. Rev. B}\ }\textbf {\bibinfo {volume} {95}},\ \bibinfo
  {pages} {165126} (\bibinfo {year} {2017})}\BibitemShut {NoStop}%
\bibitem [{\citenamefont {Cottura}\ and\ \citenamefont
  {Clouet}(2018)}]{Cottura2018}%
  \BibitemOpen
  \bibfield  {author} {\bibinfo {author} {\bibfnamefont {M.}~\bibnamefont
  {Cottura}}\ and\ \bibinfo {author} {\bibfnamefont {E.}~\bibnamefont
  {Clouet}},\ }\bibfield  {title} {\bibinfo {title} {Solubility in {Z}r-{N}b
  alloys from first-principles},\ }\href
  {https://doi.org/10.1016/j.actamat.2017.10.035} {\bibfield  {journal}
  {\bibinfo  {journal} {Acta Mater.}\ }\textbf {\bibinfo {volume} {144}},\
  \bibinfo {pages} {21} (\bibinfo {year} {2018})}\BibitemShut {NoStop}%
\bibitem [{\citenamefont {Argaman}\ \emph {et~al.}(2016)\citenamefont
  {Argaman}, \citenamefont {Eidelstein}, \citenamefont {Levy},\ and\
  \citenamefont {Makov}}]{Argaman2016}%
  \BibitemOpen
  \bibfield  {author} {\bibinfo {author} {\bibfnamefont {U.}~\bibnamefont
  {Argaman}}, \bibinfo {author} {\bibfnamefont {E.}~\bibnamefont {Eidelstein}},
  \bibinfo {author} {\bibfnamefont {O.}~\bibnamefont {Levy}},\ and\ \bibinfo
  {author} {\bibfnamefont {G.}~\bibnamefont {Makov}},\ }\bibfield  {title}
  {\bibinfo {title} {Ab initio study of the phononic origin of negative thermal
  expansion},\ }\href {https://doi.org/10.1103/physrevb.94.174305} {\bibfield
  {journal} {\bibinfo  {journal} {Phys. Rev. B}\ }\textbf {\bibinfo {volume}
  {94}},\ \bibinfo {pages} {174305} (\bibinfo {year} {2016})}\BibitemShut
  {NoStop}%
\bibitem [{\citenamefont {Poulain}(2020)}]{Poulain2020}%
  \BibitemOpen
  \bibfield  {author} {\bibinfo {author} {\bibfnamefont {R.}~\bibnamefont
  {Poulain}},\ }\emph {\bibinfo {title} {Conception et développement d’une
  nouvelle famille d’alliages de titane à haute résistance mécanique et
  biocompatibilité optimisée pour l’implantologie dentaire}},\ \href@noop
  {} {Ph.D. thesis},\ \bibinfo  {school} {Université PSL} (\bibinfo {year}
  {2020})\BibitemShut {NoStop}%
\bibitem [{\citenamefont {Toby}(2006)}]{Toby2006}%
  \BibitemOpen
  \bibfield  {author} {\bibinfo {author} {\bibfnamefont {B.~H.}\ \bibnamefont
  {Toby}},\ }\bibfield  {title} {\bibinfo {title} {${R}$ factors in {R}ietveld
  analysis: How good is good enough?},\ }\href
  {https://doi.org/10.1154/1.2179804} {\bibfield  {journal} {\bibinfo
  {journal} {Powder Diffr.}\ }\textbf {\bibinfo {volume} {21}},\ \bibinfo
  {pages} {67} (\bibinfo {year} {2006})}\BibitemShut {NoStop}%
\bibitem [{\citenamefont {Amann}\ \emph
  {et~al.}(2023{\natexlab{b}})\citenamefont {Amann}, \citenamefont {Poulain},
  \citenamefont {Delannoy}, \citenamefont {Guillou}, \citenamefont
  {Talla~Noutack}, \citenamefont {Kloenne}, \citenamefont {Nowak},
  \citenamefont {Couzinié}, \citenamefont {De~Geuser}, \citenamefont
  {Thiaudière}, \citenamefont {Béchade}, \citenamefont {Clouet},
  \citenamefont {Fraser}, \citenamefont {Guillot},\ and\ \citenamefont
  {Prima}}]{Amann2023p}%
  \BibitemOpen
  \bibfield  {author} {\bibinfo {author} {\bibfnamefont {F.}~\bibnamefont
  {Amann}}, \bibinfo {author} {\bibfnamefont {R.}~\bibnamefont {Poulain}},
  \bibinfo {author} {\bibfnamefont {S.}~\bibnamefont {Delannoy}}, \bibinfo
  {author} {\bibfnamefont {R.}~\bibnamefont {Guillou}}, \bibinfo {author}
  {\bibfnamefont {M.}~\bibnamefont {Talla~Noutack}}, \bibinfo {author}
  {\bibfnamefont {Z.}~\bibnamefont {Kloenne}}, \bibinfo {author} {\bibfnamefont
  {S.}~\bibnamefont {Nowak}}, \bibinfo {author} {\bibfnamefont {J.-P.}\
  \bibnamefont {Couzinié}}, \bibinfo {author} {\bibfnamefont {F.}~\bibnamefont
  {De~Geuser}}, \bibinfo {author} {\bibfnamefont {D.}~\bibnamefont
  {Thiaudière}}, \bibinfo {author} {\bibfnamefont {J.-L.}\ \bibnamefont
  {Béchade}}, \bibinfo {author} {\bibfnamefont {E.}~\bibnamefont {Clouet}},
  \bibinfo {author} {\bibfnamefont {H.}~\bibnamefont {Fraser}}, \bibinfo
  {author} {\bibfnamefont {I.}~\bibnamefont {Guillot}},\ and\ \bibinfo {author}
  {\bibfnamefont {F.}~\bibnamefont {Prima}},\ }\bibfield  {title} {\bibinfo
  {title} {On the oxygen ordering mechanisms in dilute {T}i-{Z}r-{O} systems},\
  }in\ \href@noop {} {\emph {\bibinfo {booktitle} {$15^{th}$ World Conference
  on Titanium}}},\ Vol.\ \bibinfo {volume} {in press}\ (\bibinfo {year}
  {2023})\BibitemShut {NoStop}%
\bibitem [{\citenamefont {Liu}\ and\ \citenamefont {Welsch}(1988)}]{Liu1988}%
  \BibitemOpen
  \bibfield  {author} {\bibinfo {author} {\bibfnamefont {Z.}~\bibnamefont
  {Liu}}\ and\ \bibinfo {author} {\bibfnamefont {G.}~\bibnamefont {Welsch}},\
  }\bibfield  {title} {\bibinfo {title} {Literature survey on diffusivities of
  oxygen, aluminum, and vanadium in alpha titanium, beta titanium, and in
  rutile},\ }\href {https://doi.org/10.1007/bf02628396} {\bibfield  {journal}
  {\bibinfo  {journal} {Metall. Trans. A}\ }\textbf {\bibinfo {volume} {19}},\
  \bibinfo {pages} {1121} (\bibinfo {year} {1988})}\BibitemShut {NoStop}%
\bibitem [{\citenamefont {Touloukian}\ \emph {et~al.}(1975)\citenamefont
  {Touloukian}, \citenamefont {Kirby}, \citenamefont {Taylor},\ and\
  \citenamefont {Desai}}]{Touloukian1975}%
  \BibitemOpen
  \bibfield  {author} {\bibinfo {author} {\bibfnamefont {Y.~S.}\ \bibnamefont
  {Touloukian}}, \bibinfo {author} {\bibfnamefont {R.~K.}\ \bibnamefont
  {Kirby}}, \bibinfo {author} {\bibfnamefont {R.}~\bibnamefont {Taylor}},\ and\
  \bibinfo {author} {\bibfnamefont {P.~D.}\ \bibnamefont {Desai}},\ }\bibfield
  {title} {\bibinfo {title} {Thermal expansion - metallic elements and
  alloys},\ }in\ \href@noop {} {\emph {\bibinfo {booktitle} {Thermophysical
  Properties of Matter Thermal Expansion}}},\ Vol.~\bibinfo {volume} {12}\
  (\bibinfo  {publisher} {Plenum},\ \bibinfo {address} {New York},\ \bibinfo
  {year} {1975})\BibitemShut {NoStop}%
\bibitem [{\citenamefont {Souvatzis}\ \emph {et~al.}(2007)\citenamefont
  {Souvatzis}, \citenamefont {Eriksson},\ and\ \citenamefont
  {Katsnelson}}]{Souvatzis2007}%
  \BibitemOpen
  \bibfield  {author} {\bibinfo {author} {\bibfnamefont {P.}~\bibnamefont
  {Souvatzis}}, \bibinfo {author} {\bibfnamefont {O.}~\bibnamefont
  {Eriksson}},\ and\ \bibinfo {author} {\bibfnamefont {M.~I.}\ \bibnamefont
  {Katsnelson}},\ }\bibfield  {title} {\bibinfo {title} {Anomalous thermal
  expansion in $\alpha$-titanium},\ }\href
  {https://doi.org/10.1103/physrevlett.99.015901} {\bibfield  {journal}
  {\bibinfo  {journal} {Phys. Rev. Lett.}\ }\textbf {\bibinfo {volume} {99}},\
  \bibinfo {pages} {015901} (\bibinfo {year} {2007})}\BibitemShut {NoStop}%
\bibitem [{\citenamefont {Mei}\ \emph {et~al.}(2009)\citenamefont {Mei},
  \citenamefont {Shang}, \citenamefont {Wang},\ and\ \citenamefont
  {Liu}}]{Mei2009}%
  \BibitemOpen
  \bibfield  {author} {\bibinfo {author} {\bibfnamefont {Z.-G.}\ \bibnamefont
  {Mei}}, \bibinfo {author} {\bibfnamefont {S.-L.}\ \bibnamefont {Shang}},
  \bibinfo {author} {\bibfnamefont {Y.}~\bibnamefont {Wang}},\ and\ \bibinfo
  {author} {\bibfnamefont {Z.-K.}\ \bibnamefont {Liu}},\ }\bibfield  {title}
  {\bibinfo {title} {Density-functional study of the thermodynamic properties
  and the pressure{\textendash}temperature phase diagram of {T}i},\ }\href
  {https://doi.org/10.1103/physrevb.80.104116} {\bibfield  {journal} {\bibinfo
  {journal} {Phys. Rev. B}\ }\textbf {\bibinfo {volume} {80}},\ \bibinfo
  {pages} {104116} (\bibinfo {year} {2009})}\BibitemShut {NoStop}%
\bibitem [{\citenamefont {Perdew}\ and\ \citenamefont
  {Zunger}(1981)}]{Perdew1981}%
  \BibitemOpen
  \bibfield  {author} {\bibinfo {author} {\bibfnamefont {J.~P.}\ \bibnamefont
  {Perdew}}\ and\ \bibinfo {author} {\bibfnamefont {A.}~\bibnamefont
  {Zunger}},\ }\bibfield  {title} {\bibinfo {title} {Self-interaction
  correction to density-functional approximations for many-electron systems},\
  }\href {https://doi.org/10.1103/physrevb.23.5048} {\bibfield  {journal}
  {\bibinfo  {journal} {Phys. Rev. B}\ }\textbf {\bibinfo {volume} {23}},\
  \bibinfo {pages} {5048} (\bibinfo {year} {1981})}\BibitemShut {NoStop}%
\end{thebibliography}%

\end{document}